\documentclass[iop,apj]{emulateapj}

\newcommand{\msun}{\ensuremath{M_{\odot}}}
\newcommand{\mjup}{\ensuremath{M_{\rm Jup}}}
\newcommand{\teff}{\ensuremath{T_{\rm eff}}}
\newcommand{\logg}{\ensuremath{\log {\rm g}}}

\newcommand{\kshort}{\ensuremath{K_{\rm s}}}

\newcommand{\lprime}{\ensuremath{L^\prime}}

\usepackage{booktabs}
\usepackage{epstopdf}
\usepackage{color}

\shorttitle{Wide substellar companions in USco}
\shortauthors{Lachapelle et al.}

\begin{document}

\title{Characterization of low-mass, wide-separation substellar companions to stars in Upper Scorpius: near-infrared photometry and spectroscopy}

\author{Fran\c cois-Ren\'e Lachapelle\altaffilmark{1}, David Lafreni\`ere\altaffilmark{1}, Jonathan Gagn\'e\altaffilmark{1}, Ray Jayawardhana\altaffilmark{2}, Markus Janson\altaffilmark{3}, Christiane Helling\altaffilmark{4}, Soeren Witte\altaffilmark{5}}
\altaffiltext{1}{Institute for Research on Exoplanets, Universit\'e de Montr\'eal, D\'epartement de Physique, C.P.~6128 Succ. Centre-ville, Montr\'eal, QC H3C~3J7, Canada.}
\altaffiltext{2}{University of Toronto, Toronto, ON, Canada}
\altaffiltext{3}{Queen's University, Belfast, Northern Ireland, UK}
\altaffiltext{4}{University of St Andrews, St Andrews, Scotland, UK}
\altaffiltext{5}{Hamburger Sternwarte, Hamburg, Germany}
\email{lachapelle@astro.umontreal.ca}

\begin{abstract}
We present new 0.9--2.45\,$\mu$m spectroscopy ($R \sim 1000$), and $Y$, $J$, $H$, $K_s$, $L^\prime$ photometry, obtained at Gemini North, of three low-mass brown dwarf companions on wide orbits around young stars of the Upper~Scorpius OB association: 
HIP~78530~B, 
[PGZ2001]~J161031.9--191305~B, and 
GSC~06214-00210~B. We use these data to assess the companions' spectral type, temperature, surface gravity and mass, as well as the ability of the {\sc BT-SETTL} and {\sc Drift-Phoenix} atmosphere models to reproduce the spectral features of young substellar objects. For completeness, we also analyze the archival spectroscopy and photometry of the Upper~Scorpius planetary mass companion 1RXS~J160929.1--210524~b.
Based on a comparison with model spectra we find that the companions, in the above order, have effective temperatures of 
2700$\pm$100\,K, 
2500$\pm$200\,K, 
2300$\pm$100\,K and 
1700$\pm$100\,K. These temperatures are consistent with our inferred spectral types, respectively 
M7\,$\beta$, 
M9\,$\gamma$, 
M9\,$\gamma$, and 
L4\,$\gamma$, obtained from spectral indices and comparisons with templates.
From bolometric luminosities estimated from atmosphere model spectra adjusted to our photometry, and using evolution models at 5--10~Myr, we estimate masses of 21--25\,\mjup, 28--70\,\mjup, 14--17\,\mjup, and 7--12\,\mjup, respectively. 
[PGZ2001]~J161031.9--191305~B appears significantly over-luminous for its inferred temperature, which explains its higher mass estimate.
Synthetic spectra based on the {\sc BT-Settl} and {\sc Drift-Phoenix} atmosphere models generally offer a good fit to our observed spectra, although our analysis has highlighted a few problems.
For example, the best fits in the individual near-infrared bands occur at different model temperatures.
Also, temperature estimates based on a comparison of the broadband magnitudes and colors of the companions to synthetic magnitudes from the models are systematically lower than the temperature estimates based on a comparison with synthetic spectra.
\end{abstract}


\section{INTRODUCTION} 

Since the first confirmed detection of a brown dwarf in 1995 \citep{1995Sci...270.1478O}, a substantial effort has been made on characterizing substellar objects. Up to now, almost 2000 isolated brown dwarfs have been discovered, and we are beginning to better understand their formation and evolution. Nevertheless, the modeling of their cool atmospheres, bearing several molecules and dust clouds, is a great challenge for modern astrophysics. 
Several low-mass substellar companions ($\lesssim$\,30\,\mjup) have been discovered recently on wide orbits ($>$\,80\,AU) around stars, see for example \citet{Neuhauser:2012ur} and references therein. 
The age and distance of these companions can be inferred from their primary star, while their large separation permits their direct observation without the hampering glare of their primary star; their characterization is thus particularly interesting for testing low temperature atmosphere and evolution models. 
At young ages these companions are even more interesting as this is where an empirical verification of the theoretical models is most needed \citep{2012EAS....57....3A}. Furthermore, these young companions are potentially (more massive) analogs to the young giant planets recently imaged \citep{2008Sci...322.1348M,2008ApJ...689L.153L,2010Sci...329...57L,Carson:2013fw,Kuzuhara:2013jz,Rameau:2013dr}, and thus can serve as workbenches in support of the more difficult direct imaging studies of exoplanets. 

In this paper we present and analyze new near-infrared photometric and spectroscopic observations of a sample of three wide substellar companions to young stars in the Upper~Scorpius OB (USco) formation region: the $\sim$\,16\,\mjup\ companion at a separation of $\sim$\,320\,AU around the K7 star GSC~06214-00210 (hereafter G06214; \citealt{2011ApJ...726..113I}), the $\sim$\,23\,\mjup\ companion at 740\,AU from the B9 star HIP~78530 \citep{2011ApJ...730...42L}, and the $\sim$\,34\,\mjup\ companion at $\sim$\,885\,AU from the K7 star [PGZ2001]~J161031.9--191305 (hereafter J1610--1913; \citealt{2008ApJ...679..762K}). The primary of J1610--1913 is itself a tight binary (Aab, separation of $\sim$\,0\farcs145 or $\sim$\,20\,AU, \citealp{2008ApJ...679..762K}), in which the companion (Ab) is roughly at the stellar/substellar boundary. 
We also apply the same analysis, using existing data, to the $\sim$\,8\,\mjup\ companion at 330\,AU around the K7 star 1RXS~J160929.1--210524 (hereafter J1609--2105) that was first identified in \citet{2008ApJ...689L.153L} and further analyzed in \citet{2010ApJ...719..497L}. 
USco is located at a distance of $145\pm14$\,pc \citep{1999AJ....117..354D,Preibisch:1999jk} and the average age in the region is estimated at 5\,Myr, with a very small scatter ($\pm1$\,Myr) \citep{DeGeus:1992ww,2002AJ....124..404P}. 
It is thus reasonable to consider a common age for the systems in our sample, meaning that on a comparative basis, the companions should not be affected by the age-mass degeneracy inherent to substellar objects. 
The initial age estimate of 5\,Myr for USco was recently revised to 11$\pm$2\,Myr by \citet{Pecaut:2012ux}, based on isochrone fitting. This new age seems to be consistent with the results of \citet{2012AJ....144....8S}, given recent the revision of the age of the Beta Pictoris moving group proposed by \citet{Binks:2013gd}. Still, the exact age of USco is still the subject of debate, and is beyond the scope of this paper, so in our work below we present results assuming both estimates.

The four companions studied in this paper have all been studied to various degrees in earlier publications. The near-infrared photometry and spectroscopy of HIP~78530~B was analyzed in \citet{2011ApJ...730...42L}. \citet{2013ApJ...767...31B} studied HIP~78530~B, G06214~B, and J1609--2105~b using 3--5\,$\mu$m photometry. \citet{Bowler:2011gw} presented 1.1--1.8\,$\mu$m spectroscopy of G06214~B, adding to the $JHKL^\prime$ photometry from \citet{2011ApJ...726..113I}. The latter study also independently confirmed the common proper motion of J1609--2105~b. 
J1610--1913~B was observed in the $K_S$ band by \citet{2008ApJ...679..762K}, who made its discovery. Recently, \citet{2013ApJ...773...63A} presented a low resolution ($R \sim 100$) 0.8--2.5\,$\mu$m spectrum and $H$- and $K$-band photometric measurements of J1610--1913~B. 
In this paper, in addition to carrying out a homogeneous analysis of the four companions, with a focus on a comparison of their spectra with those of atmosphere models, we also present new data. We present the first photometric measurements in $Y$, $J$ and $L^\prime$ of J1610--1913~B. For HIP~78530~B, the $Y$-band photometry and the 0.90--1.15\,$\mu$m spectrum have never been published before. The spectrum of G06214~B in the $K$ band is also presented for the first time, along with the part of the $J$ band between 1.00\,$\mu$m and 1.18\,$\mu$m, where the VO band and the \ion{Na}{1} and the two first \ion{K}{1} lines are found. We also present the first $Y$-band photometric measurement for G06214~B.

\section{OBSERVATIONS AND DATA REDUCTION} 

\begin{centering}
\begin{deluxetable*}{llllllll}
\tablewidth{0pt}
\tablecolumns{8}
\tablecaption{\label{tbl:obs} Observation log}
\tablehead{\colhead{Target} &\colhead{Date} & \colhead{Mode} & \multicolumn{5}{c}{Total Integration Time Per Filter (s)}\\
&&&\multicolumn{5}{c}{$(t_{\rm non-saturated}$, $t_{\rm saturated}$)}
}\startdata
HIP~78530 & 2011-03-30 & Imaging & $Y$(25, 50), &&&$K_{\rm con.}$(30, 30) &\\
J1610--1913 & 2011-04-19 & Imaging & $Y$(175, \--), &$J$(75, \--), &$H$(50, \--), &$K_s$(45, 50)&\\
G06214 & 2011-04-19 & Imaging & $Y$(40, 50), &$J$(30, 50), &$H$(50, \--), &$K_s$(60, \--)&\\
J1610--1913 & 2011-06-20 & Spectro & \multicolumn{4}{c}{\--- (600, \--)}& \\
G06214 & 2011-06-23 & Spectro & \multicolumn{4}{c}{\--- (2880, \--)}& \\
HIP~78530 & 2011-07-03 & Spectro & \multicolumn{4}{c}{\--- (1350, \--)}& \\
HIP~78530 & 2011-08-16 & Imaging & &&&&$L^\prime$(288, \--) \\
J1610--1913 & 2011-08-16 & Imaging & &&&&$L^\prime$(90, \--)\\
\enddata
\end{deluxetable*}
\end{centering}

\subsection{Imaging}

The imaging observations were performed at the Gemini North telescope in semester 2011A (program GN-2011A-Q-60) using the NIRI camera in combination with the ALTAIR adaptive optic (AO) system \citep{2000SPIE.4007..115H}. The primary stars themselves were used for wavefront sensing and the ALTAIR field lens was used to reduce the effects of anisoplanatism and achieve better image quality at separations of a few arc seconds. The $f/32$ camera was used, resulting in a pixel scale of $0.0214\arcsec$ and a field of view of $22\arcsec \times 22\arcsec$.\footnote{As given on the instrument web page at \url{http://www.gemini.edu/sciops/instruments/niri/imaging/pixel-scales-and-fov}.} The Cassegrain rotator was turned off during the observations, to match the setup used for earlier observations of the same stars, and thus the field of view orientation changed slowly during the sequences.
For HIP~78530, we took observations with the $Y$ filter, to complement similar observations made previously in $J$, $H$ and \kshort\ and initially reported in \citet{2011ApJ...730...42L}. We also obtained observations of HIP~78530 in the narrowband $K_{\rm cont}$ filter (2.0975\,$\mu$m) for astrometric follow up as observations at three earlier epochs had already been obtained in this filter.
For GSC~06214 and J1610--1913, we took images with the $Y$, $J$, $H$, and \kshort\ filters. 
The observation log is presented in table~\ref{tbl:obs}.

For all targets we used a pattern of five dither positions consisting of the centre and corners of a square of $10\arcsec$ on one side. For most observations, the primary is too bright to get a good signal from the companion without saturating the detector. To obtain deeper images allowing more precise photometry of the companions, we thus obtained, at each dither position, a set comprising unsaturated images consisting of multiple co-additions of short integrations in fast, high read-noise mode, followed by one saturated image consisting of one long integration in slow, low read-noise mode. The saturated images can be readily registered to the unsaturated images and easily corrected in the saturated part using the properly scaled unsaturated images.

We also observed HIP~78530 and J1610--1913 in \lprime , still with the $f/32$ camera but without the use of AO. We observed the faint photometric standard star FS~140 \citep{Leggett:2003gt} shortly after the targets to calibrate the \lprime\ photometric measurements. For these observations, we used a pattern of five $A-B$ nod pairs with a separation of 8\arcsec, each pair being displaced from the preceding one by 2\arcsec.
At each position we obtained 12 co-additions with an integration time of 0.75~s each, ensuring that the primary star was never saturated. This sequence was repeated three times for HIP~78530, with a pattern rotation of $90\degr$ between each sequence. For J1610--1913, the sequence was executed only once.

We reduced the data using custom {\tt IDL} routines.
For the images obtained in the high read-noise mode, a striped noise pattern was often present and we removed it by proper median filtering.
For the images taken in $Y$, $J$, $H$, and $K$, we constructed a sky frame by taking the median of the images at all dither positions, after masking out the sources in each one. For the images in \lprime, we built the sky frame as the mean of the two images obtained at the preceding and following dither positions (also after masking out the sources). After subtraction of the sky frame, we divided the images by a normalized flat field, and we corrected the geometric distortion of the images using the prescription given on the instrument webpage.\footnote{The  distortion is given by $r^\prime = r + k*r^2$, where $k = (1.32 \pm 0.02) \times 10^{-5}$, $r$ is the uncorrected distance from the field centre in pixels, and $r^\prime$ is the corrected distance from the centre in pixels. From \url{http://www.gemini.edu/sciops/instruments/niri/imaging/pixel-scales-and-fov}.} The reduced images at each dither position were then registered to place the primary star at their centre, de-rotated to a common field orientation, and their median was taken. The saturated region of the long-exposure images were finally replaced by the properly scaled unsaturated images. For improved consistency among all of our targets, we reprocessed the archival $J$-, $H$- and $K$-band data for HIP~78530~B \citep{2011ApJ...730...42L}.

\subsection{Spectroscopy}

The spectroscopy observations were made at the Gemini North telescope in the same program as the imaging, using the GNIRS spectrograph \citep{Elias:2006dh} in cross-dispersed (XD) spectroscopy mode with a $0\farcs45$-wide slit, the 10 lines mm$^{-1}$ grating and the long blue camera with its LXD prism, resulting in a coverage from 0.885\,$\mu$m to 2.425\,$\mu$m, see observation log in table~\ref{tbl:obs}. The ALTAIR AO system was also used to improve the spatial resolution and image quality, and thus greatly reduce the contamination from the bright nearby primary at the position of the companion. Given the wide slit used, the spectral resolving power achieved is determined by the width of the AO-corrected point-spread function (PSF) ($\sim$\,140--190\,mas) and varied between 900 and 1300 depending on target and wavelength. We obtained three exposures of 100 to 360 s integration, depending on the source, at each of two nod positions along the slit (for sky subtraction). We observed the A0 telluric standard star~HD~151787 \citep{Houk:1988wv} immediately after each target to determine and correct for the effect of the atmospheric and instrumental transmissions. Observatory standard calibration data (flat field, arc lamps) were obtained with each observation.

We reduced the data using custom {\tt IDL} routines. First, we subtracted the exposures taken at two different nods in the slit to remove the majority of the background signal, resulting in parallel positive and negative signal traces. We then divided the frames by a normalized flat field, using a different lamp in the $K$ band than for the rest of the spectrum for saturation considerations. A few frames also presented a noise pattern of stripes that was removed by carefully applying 1D iterative median filtering for each quadrant separately.
We then rectified the traces of each order using cubic interpolation. We next corrected a slight spectral shearing by rectifying atmospheric lines that were apparent on the frames before subtracting the two nod positions. With AO, the shape of the PSF is wavelength dependent. 
We thus fitted the trace independently for each spectral pixel along the spectrum. We fitted an analytic trace consisting of the sum of a Gaussian profile for its core and a Moffat profile for the wings. While fitting for the trace, we simultaneously fitted (and removed) the potential contribution from the primary star and any residual background signal.
The contamination from the primary depends on the contrast and separation of the companion and is most important for G06214~B, with only $\sim$\,2\,\arcsec\ separation. 
The flux from each nod position, cleaned from contamination and residual background signal, was extracted separately using the fitted trace as weight; the flux from the two positions were then summed together. 
The wavelengths were calibrated using an Ar arc lamp exposure and the different orders of the spectrum were combined by adjusting their overlapping sections.
Then, we divided the target spectra by the total transmission function. The latter was determined from the spectrum of the telluric standard. Namely, the continuum of the standard star spectrum was modelled by a black body function and removed, while its hydrogen absorption lines were fitted by a Voigt profile over the appropriate wavelength ranges, and then divided out.
The median of all individual spectra was taken as the final spectrum and their dispersion was used to estimate the uncertainties.
Based on the achieved PSF FWHM, the effective resolving power in the $H$ band for the different spectra were $R \sim 1110$ for HIP~78530~B, $R \sim 930$ for G06214~B and $R \sim 1260$ for J1610--1913~B.

\section{ANALYSIS AND RESULTS}

\subsection{Photometry and astrometry}\label{sect:phot-astrom}

For each system, the position of the primary and companion was measured by fitting an elongated 2D gaussian function to their PSF. For saturated PSFs, the position from the preceding unsaturated frame was used. The flux ratio between the companion and primary was calculated using aperture photometry, with an aperture radius set to the radius at which the radial intensity profile of the companion falls below the $1\sigma$ background noise level. The contribution of the primary star flux inside the photometry aperture of the companion was estimated, and removed, in the following manner. First, an azimuthally symmetric median radial intensity profile of the central star was calculated and subtracted from the image. Then a similar profile was calculated for the companion in the residual image, and this profile was subtracted from the original image. This process was then repeated once to ensure that the radial profile of the primary was not biased by the companion. The flux measurement for the companion was performed on the original image to which we subtracted the modelled flux of the primary star, while the measurement for the primary was made on the original image minus the modelled flux of the companion. These measurements were performed on the combined images as well as on the individually reduced frames. The uncertainties on the separation, position angle and photometry of the companions were determined from the scatter of measurements from individual frames. The pixel scale, $(0\farcs0214$~pixel$^{-1})$, was taken from the instrument manual and the direction towards North was taken from the image headers. By comparing our 2011 measurements with measurements of the same systems made at earlier epochs, we noticed that our position angle values were systematically off by $(-0.45\pm0.04)\degr$, based on previous measurements on 3 targets; we thus corrected our measurements for this systematic offset and included it in the position angle errors. The flux ratios between the primaries and companions are given in table~\ref{tbl:deltaphot}, along with their angular separations and position angles.

\begin{centering}
\begin{deluxetable*}{lcccc}
\tablewidth{0pt}
\tablecolumns{5}
\tablecaption{\label{tbl:deltaphot} Measured astrometric and photometric parameters}
\tablehead{\colhead{} &\colhead{HIP~78530~B} & \colhead{J1610--1913~Ab} & \colhead{J1610--1913~B} & \colhead{G06214~B}}\startdata
Angular separation (\arcsec) 	&	$4.527\pm0.003$		&	$0.171\pm0.002$	&	$5.943\pm0.002$	&	$2.204\pm0.002$ \\
Position angle (deg)\tablenotemark{a} 	&	$140.30\pm0.1$		&	$90.6\pm0.4$	&	$113.77\pm0.08$	&	$175.97\pm0.05$ \\
$\Delta Y$ (mag)         	&	$9.5\pm0.3$		&	$2.78\pm0.03$	&	$4.64\pm0.03$	&	$7.00\pm0.26$ \\
$\Delta J$ (mag)         	&	 $8.28\pm0.05$\tablenotemark{b}		&	$2.54\pm0.06$	&	$4.02\pm0.02$	&	$6.18\pm0.03$ \\
$\Delta H$ (mag)         	&	 $7.61\pm0.03$\tablenotemark{b}		&	$2.45\pm0.02$	&	$4.11\pm0.02$	&	$6.19\pm0.02$ \\
$\Delta K_{\rm s}$ (mag) 	&	$7.28\pm0.03$\tablenotemark{b}		&	$2.51\pm0.06$	&	$3.85\pm0.02$  &	$5.74\pm0.01$\\
$\Delta K^{\rm 2.09}_{\rm cont}$ (mag) 	&	$7.27\pm0.07$&	\nodata	&	\nodata	&	\nodata		\\
$\Delta L^\prime$ (mag)         	&	$6.9\pm0.2$		&	$2.50\pm0.05$	&	$3.33\pm0.04$ 	&	$4.75\pm0.05$\tablenotemark{c}\\
\enddata
\tablenotetext{a}{Corrected for a  $-0.45\pm0.04\degr$ offset.}
\tablenotetext{b}{Remeasured from observations of \citet{2011ApJ...730...42L}.}
\tablenotetext{c}{From \citet{2011ApJ...726..113I}.}
\end{deluxetable*}
\end{centering}

\begin{centering}
\begin{deluxetable}{lcc}
\tablewidth{\linewidth}
\tablecolumns{3}
\tablecaption{\label{tbl:78530} Properties of HIP~78530}
\tablehead{
\colhead{} & \multicolumn{2}{c}{Value} \\
\cline{2-3}
\colhead{Parameter} & \colhead{Primary} & \colhead{Companion}}
\startdata
$Y$ (mag)       	&	 $6.766\pm0.020$\tablenotemark{a} 	&	 $16.27\pm0.05$ \\
$J$ (mag)       	&	 $6.925\pm0.021$\tablenotemark{b}  	&	 $15.21\pm0.05$ \\
$H$ (mag)      	&	 $6.931\pm0.029$\tablenotemark{b}  	&	 $14.55\pm0.04$ \\
$K_{\rm s}$ (mag)	&	 $6.900\pm0.020$\tablenotemark{b} 	&	 $14.18\pm0.04$ \\
$L^\prime$ (mag)	&	 $6.91\pm0.02$ 	&	 $13.81\pm0.20$ \\
$J-K_{\rm s}$ (mag)\tablenotemark{c} 	&	 $-0.57\pm0.03$ 	&	 $0.95\pm0.06$ \\
$H-K_{\rm s}$ (mag)\tablenotemark{c} 	&	 $0.00\pm0.04$ 	&	 $0.34\pm0.06$ \\
$K_{\rm s}-L^\prime$ (mag)\tablenotemark{c} 	&	 $-0.04\pm0.03$ 	&	 $0.3\pm0.2$ \\
Spectral type 	&	 B9V\tablenotemark{d} 	&	 M7$\pm0.5$\,$\beta$ \\
$T_{\rm eff}$ (K) 	&	 $\sim$\,10\,500\tablenotemark{e}	&	 2700$\pm100$ \\
Distance (pc)	&	 $156.7\pm13.0$\tablenotemark{f} 	&	 \nodata \\
Projected separation (AU) 	&	 \multicolumn{2}{c}{$740\pm 60$}\\		
$\log{(L/L_\odot)}$ 	&	  	&	 $-2.53\pm0.09$ \\
Mass ($M_\odot$) [5 Myr]	&	 $\sim$\,2.5\tablenotemark{e}&	 $0.022\pm 0.001$\\
Mass ($M_\odot$) [10 Myr]	&	 $\sim$\,2.5\tablenotemark{e}&	 $0.023\pm 0.002$\\
\enddata
\tablenotetext{a}{Extrapolated from a template spectrum \citep{1998PASP..110..863P} scaled to the measured flux in other bands, see text for detail.}
\tablenotetext{b}{From \emph{2MASS} PSC \citep{2006Skrutskie}, converted to the MKO system with the equations in \citet{2001AJ....121.2851C}.}
\tablenotetext{c}{Dereddened colors, see text for detail.}
\tablenotetext{d}{From \cite{Houk:1988wv}.}
\tablenotetext{e}{From \citet{2011ApJ...730...42L} and references therein.}
\tablenotetext{f}{From \citet{vanLeeuwen:2007dc}.}
\end{deluxetable}
\end{centering}

\begin{centering}
\begin{deluxetable}{lccc}
\tablewidth{\linewidth}
\tablecolumns{4}
\tablecaption{\label{tbl:1610} Properties of [PGZ2001] J161031.9--191305}
\tablehead{
\colhead{} & \multicolumn{3}{c}{Value} \\
\cline{2-4}
\colhead{Parameter} & \colhead{Primary} & \colhead{Secondary} & \colhead{Tertiary}}
\startdata
$Y$ (mag)       	&	 $10.274\pm0.020$\tablenotemark{a}   	&	 $12.65\pm0.05$ 	&	 $14.73\pm0.05$ \\
$J$ (mag)       	&	 $10.062\pm0.026$\tablenotemark{b}  	&	 $12.61\pm0.05$ 	&	 $14.09\pm0.05$ \\
$H$ (mag)        	&	 $9.337\pm0.022$\tablenotemark{b}  	&	 $11.80\pm0.04$ 	&	 $13.43\pm0.04$ \\
$K_{\rm s}$ (mag) 	&	 $9.068\pm0.021$\tablenotemark{b}  	&	 $11.58\pm0.04$	&	 $12.92\pm0.04$ \\
$L^\prime$ (mag)         	&	 $8.72\pm0.07$	&	 $11.22\pm0.07$ 	&	 $12.05\pm0.06$ \\
$J-K_{\rm s}$ (mag) \tablenotemark{c} 	&	 $0.81\pm0.03$ 	&	 $0.84\pm0.06$ 	&	 $0.98\pm0.06$ \\	
$H-K_{\rm s}$ (mag) \tablenotemark{c} 	&	 $0.20\pm0.03$ 	&	 $0.15\pm0.06$ 	&	 $0.47\pm0.06$ \\	
$K_{\rm s}-L^\prime$ (mag) \tablenotemark{c} 	&	 $0.27\pm0.07$ 	&	 $0.29\pm0.08$	&	 $0.80\pm0.07$ \\	
Spectral type 	&	K7 \tablenotemark{d}	&	$\sim$\,M4 	&	 M9$\pm0.5$\,$\gamma$ \\
$T_{\rm eff}$ (K) 	&	 $\sim$\,4000	&	 3200$\pm300$ 	&	 2500$\pm200$ \\
Distance (pc) 	&	 $145\pm14$ \tablenotemark{e} 	&	 \nodata 	&	 \nodata \\
Proj. sep. (AU) 	&	 \nodata 	&	$26\pm3$	&	$885\pm85$ \\
$\log{(L/L_\odot)}$ 	&	 \nodata 	&	 $-1.48\pm0.11$ 	&	 $-2.13\pm0.12$ \\	
Mass ($M_\odot$)  [5 Myr]	&	 $\sim$\,0.77 \tablenotemark{f}	&	 $0.12\pm0.02$ 	&	 $0.032\pm0.004$\\
Mass ($M_\odot$)  [10 Myr]	&	 $\sim$\,0.77 \tablenotemark{f}	&	 $0.16\pm0.02$ 	&	 $0.058\pm0.011$\\
\enddata
\tablenotetext{a}{Extrapolated from a template spectrum \citep{1998PASP..110..863P} scaled to the measured flux in other bands, see text for detail.}
\tablenotetext{b}{Resolved MKO photometry based on our measured contrast and unresolved \emph{2MASS} PSC photometry \citep{2006Skrutskie}, table~\ref{tbl:deltaphot}, using the system conversion equations in \citet{2001AJ....121.2851C}.}
\tablenotetext{c}{Dereddened colors, see text for detail.}
\tablenotetext{d}{From \citet{Preibisch:2001ey}.}
\tablenotetext{e}{Mean distance of USco from \citet{1999AJ....117..354D}, with uncertainties discussed in \citet{2011ApJ...726..113I}.}
\tablenotetext{f}{From \citet{2008ApJ...679..762K}.}
\end{deluxetable}
\end{centering}

\begin{centering}
\begin{deluxetable}{lcc}
\tablewidth{\linewidth}
\tablecolumns{3}
\tablecaption{\label{tbl:06214} Properties of GSC 06214-00210}
\tablehead{
\colhead{} & \multicolumn{2}{c}{Value} \\
\cline{2-3}
\colhead{Parameter} & \colhead{Primary} & \colhead{Companion}}
\startdata
$Y$ (mag)       	&	 $10.20\pm0.020$\tablenotemark{a} 	&	 $17.20\pm0.05$ \\
$J$ (mag)     	&	 $9.946\pm0.027$\tablenotemark{b} 	&	 $16.13\pm0.04$ \\
$H$ (mag)	&	 $9.329\pm0.024$\tablenotemark{b} 	&	 $15.52\pm0.03$ \\
$K_{\rm s}$ (mag)	&	 $9.129\pm0.021$\tablenotemark{b} 	&	 $14.87\pm0.02$ \\
$L^\prime$ (mag)	&	 $9.10\pm0.05$ 	&	 $13.75\pm0.07$ \\
$J-K_{\rm s}$ (mag)\tablenotemark{c} 	&	 $0.72\pm0.03$ 	&	 $1.16\pm0.04$ \\
$H-K_{\rm s}$ (mag)\tablenotemark{c} 	&	 $0.63\pm0.04$ 	&	 $0.61\pm0.04$ \\
$K_{\rm s}-L^\prime$ (mag)\tablenotemark{c} 	&	 $-0.01\pm0.05$ 	&	 $1.08\pm0.07$ \\
Spectral type 	&	K7$\pm0.5$\tablenotemark{d} 	&	  M9$\pm0.5$\,$\gamma$ \\
$T_{\rm eff}$ (K) 	&	 $4200\pm150$\tablenotemark{d} 	&	 2300$\pm100$ \\
Distance (pc) 	&	 $145\pm14$\tablenotemark{e} 	&	 \nodata \\
Projected separation (AU) 	&	 \multicolumn{2}{c}{$320\pm30$} \\
$\log{(L/L_\odot)}$ 	&	$-0.42\pm0.08$\tablenotemark{d}  	&	 $-3.01\pm0.09$ \\
Mass ($M_\odot$) [5 Myr]	&	 $0.9\pm0.1$\tablenotemark{d} 	&	 $0.015\pm0.001$\\
Mass ($M_\odot$) [10 Myr] 	&	 $0.9\pm0.1$\tablenotemark{d} 	&	 $0.016\pm0.001$\\
\enddata
\tablenotetext{a}{Extrapolated from a template spectrum \citep{1998PASP..110..863P} scaled to the measured flux in other bands, see text for detail.}
\tablenotetext{b}{From \emph{2MASS} PSC \citep{2006Skrutskie}, converted in MKO with equations in \citet{2001AJ....121.2851C}.}
\tablenotetext{c}{Dereddened colors, see text for detail.}
\tablenotetext{d}{From \citet{Bowler:2011gw}.}
\tablenotetext{e}{Mean distance of USco from \citet{1999AJ....117..354D}, with uncertainties discussed in \citet{2011ApJ...726..113I}.}
\end{deluxetable}
\end{centering}

\begin{centering}
\begin{deluxetable}{lcc}
\tablewidth{0pt}
\tablecolumns{3}
\tablecaption{\label{tbl:1609} Properties of 1RXS~J160929.1--210525}
\tablehead{
\colhead{} & \multicolumn{2}{c}{Value} \\
\cline{2-3}
\colhead{Parameter} & \colhead{Primary} & \colhead{Companion}}
\startdata
$J$ (mag)     	&	 $9.764\pm0.027$\tablenotemark{a} 	&	 $17.85\pm0.12$\tablenotemark{b} \\
$H$ (mag)	&	 $9.109\pm0.023$\tablenotemark{a} 	&	 $16.86\pm0.07$\tablenotemark{b} \\
$K_{\rm s}$ (mag)	&	 $8.891\pm0.021$\tablenotemark{a} 	&	 $16.15\pm0.05$\tablenotemark{b} \\
$L^\prime$ (mag)\tablenotemark{c}	&	 $8.73\pm0.05$ 	&	 $14.8\pm0.3$ \\
$J-K_{\rm s}$ (mag) 	&	 $0.87\pm0.03$ 	&	 $1.7\pm0.1$ \\
$H-K_{\rm s}$ (mag) 	&	 $0.22\pm0.04$ 	&	 $0.71\pm0.09$ \\
$K_{\rm s}-L^\prime$ (mag) 	&	 $0.16\pm0.05$ 	&	 $1.4\pm0.3$ \\
Spectral type 	&	K7~V$\pm1$\tablenotemark{c} 	&	  L4$\pm1$\,$\gamma$ \\
$T_{\rm eff}$ (K) 	&	 $4060^{+300}_{-200}$\tablenotemark{c} 	&	 $1700\pm100$ \\
Distance (pc) 	&	 $145\pm14$\tablenotemark{d} 	&	 \nodata \\
Projected separation (AU) 	&	 \multicolumn{2}{c}{$\sim 330$} \\
$\log{(L/L_\odot)}$ 	&	$-0.37\pm0.15$\tablenotemark{c}  	&	 $-3.5\pm0.2$ \\
Mass ($M_\odot$) [5 Myr]	&	 $0.85^{+0.20}_{-0.10}$\tablenotemark{c} 	&	 $0.008\pm0.001$\\
Mass ($M_\odot$) [10 Myr]	&	 $0.85^{+0.20}_{-0.10}$\tablenotemark{c} 	&	 $0.011\pm0.001$\\
\enddata
\tablenotetext{a}{From \emph{2MASS} PSC \citep{2006Skrutskie}, converted to the MKO system with the equations in \citet{2001AJ....121.2851C}.}
\tablenotetext{b}{Based on the contrasts given in \citet{2008ApJ...689L.153L}.}
\tablenotetext{c}{From \citet{2008ApJ...689L.153L}.}
\tablenotetext{d}{Mean distance of USco from \citet{1999AJ....117..354D}, with uncertainties discussed in \citet{2011ApJ...726..113I}.}
\end{deluxetable}
\end{centering}

The companion apparent magnitudes were computed from the measured contrast ratios in combination with the $J$, $H$ and $K_{\rm s}$ magnitudes of the primaries taken from \emph{2MASS} \citep{2006Skrutskie}. 
For the J1610--1913 system, the photometric measurements from \emph{2MASS} did not resolve the tight binary. We thus corrected the photometry of the primary according to our measured contrast for the tight binary.
The \emph{2MASS} magnitudes were converted to the MKO system using the transformation equations from the online supplements\footnote{\url{http://www.astro.caltech.edu/$\sim$jmc/2mass/v3/transformations/}.} of \citet{2001AJ....121.2851C}. 
The $Y$-band magnitudes of the primaries were estimated from template spectra of the appropriate spectral type taken from the Pickles Atlas stellar spectral flux library\footnote{\url{http://www.stsci.edu/hst/observatory/cdbs/pickles\_atlas.html}.} \citep{1998PASP..110..863P}. 
Using the MKO filter profiles and zero points from \citet{2005PASP..117.1459T} and online supplements\footnote{\url{http://irtfweb.ifa.hawaii.edu/IRrefdata/iwafdv.html}.}, the atlas spectra were scaled to fit our measured fluxes in the $JHK_{\rm s}$ bands, and then integrated over the $Y$ filter to get the synthetic $Y$-band flux, and thus the $Y$-band magnitude of the star.
For the $L^\prime$ band, we measured the magnitudes of the primaries and companions directly from our images, using our observations of the faint photometric standard star FS~140 \citep{Leggett:2003gt} for calibration, as mentioned earlier. 
Tables~\ref{tbl:78530}, \ref{tbl:1610}, \ref{tbl:06214}, and \ref{tbl:1609} present the resulting photometry and colors for each system. 
The colors in these tables and in Figure~\ref{fig:colcol}, along with the spectra presented below, have been corrected for interstellar extinction using the $YJHK$-band absorption coefficients from \citet{Cardelli:1989dp} and the $A(L)/A(V)$ ratio from \citet{Allen:2000tl}, assuming $R_V=3.1$. \citet{Carpenter:2009bg} published extinction values of $A_V=0.5$ for HIP~78530, $A_V=1.1$ for J1610--1913 and $A_V=0.0$, for J1609--2105. For G06214, \citet{2013ApJ...767...31B} found that its extinction is consistent with $A_V=0.0$, which we adopted. We also took the extinction into account when measuring the $Y$-band photometry using the procedure described above, by fitting extinction-corrected fluxes to the template spectra, calculating the synthetic $Y$-band magnitude, and then applying the proper correction to our measurement.

The new $K_{\rm cont}$ data for HIP~78530~B were used to further assess the common proper motion of this companion with its primary, as some doubts about its physical association were raised by \citet{2013ApJ...767...31B}, who mentioned that it could possibly be an early-M background star. At epoch 2011.2422, we measure a separation and position angle of $4.527\arcsec\pm0.003\arcsec$ and $140.30\degr\pm0.1\degr$, respectively, which are consistent with the values of $4.529\arcsec\pm0.006\arcsec$ and $140.32\degr\pm0.1\degr$ measured by \citet{2011ApJ...730...42L} for epoch 2008.3940. Over that time, the separation and position angle of a (stationary) background star would have decreased by $0.034\arcsec$ and $0.82\degr$, respectively. Thus our new measurements indicate with increased significance ($\sim$\,10$\sigma$) that the companion is co-moving with the primary. Moreover, the new spectrum we have acquired is inconsistent with the companion being an early-M background star (see Sections \ref{sec:spec} and \ref{sec:D78530}). 

\subsection{Spectroscopy}\label{sec:spec}

\begin{figure*}
\plotone{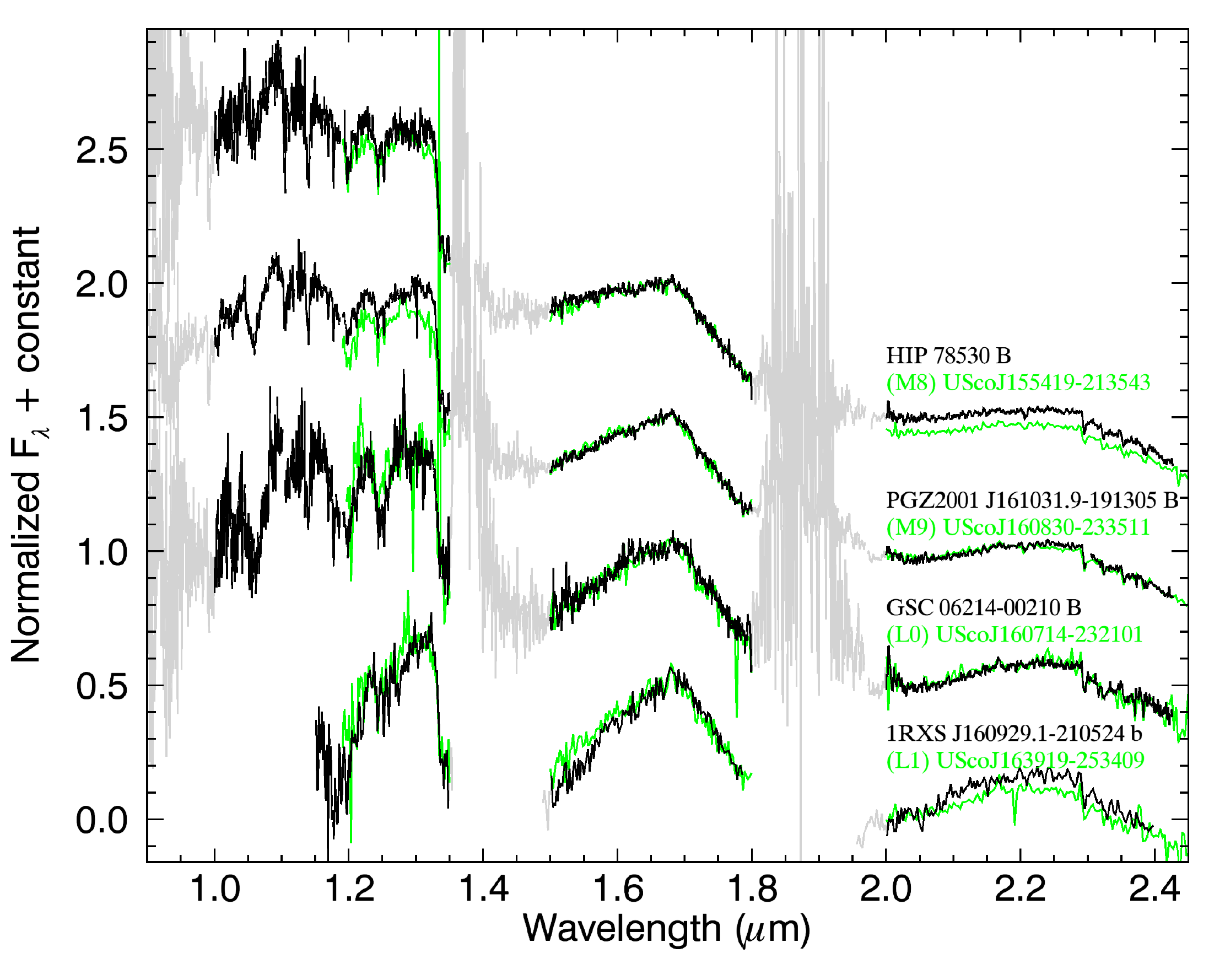}
\caption{\label{fig:compare} In black from top to bottom, our GNIRS-XD spectra of HIP~78530~B, J1610--1913~B, and G06214~B, and the archival spectrum of J1609--2105~b from \citet{2010ApJ...719..497L}. The spectra of HIP~78530~B and J1610--1913~B have been corrected for extinction (see text). Regions of strong telluric absorption have been greyed out.
In green from top to bottom, comparison spectra of USco brown dwarfs from \citet{Lodieu:2007hm}: USco~J155419--213543 (M8), USco~J160830--233511 (M9), USco~J160714--232101 (L0), and USco~J163919--253409 (L1).
}
\end{figure*}

The newly obtained spectra of HIP~78530~B, G06214~B, and J1610--1913~B are shown in Figure~\ref{fig:compare}. The spectrum of J1609--2105~b from \citet{2010ApJ...719..497L} is also shown.
The average per pixel signal-to-noise (S/N) ratios of our three spectra over the whole spectral range are $\sim$\,130 for J1610--1913~B, $\sim$\,60 for HIP~78530~B  and $\sim$\,30 for G06214~B. The lower S/N of the latter is due to its lower brightness, combined with a relatively more important contamination from the primary. The spectra display the typical morphologies of young late-M dwarfs, with prominent water absorption bands. The spectra of the four objects also show a smooth gradation in all three spectral bands. From top (HIP~78530~B) to bottom (J1609--2105~b) in the figure, the $J$-band spectrum shows increasingly deeper VO and FeH absorption bands. Furthermore, the slopes of the blue side of both the $H$ and $K$ bands become increasingly more pronounced, owing to stronger absorption by water vapor.

We used the method of K.~Cruz et al. (in preparation; see \citealt{DisentanglingLDwar:db}) to assign spectral types to the objects presented here. The method consists of a band-per-band visual comparison with field, intermediate-gravity and very low-gravity spectroscopic templates that were constructed from a median combination of several spectra that were assigned the same spectral type and gravity class in the NIR. We verified that the classifications of \citet{Lodieu:2007hm} and \citet{2013ApJ...772...79A}, based on the H$_2$O index from \citet{2007ApJ...657..511A} and the H$_2$O-1, H$_2$O-2 and FeH indices from \citet{2004ApJ...610.1045S}, generally agreed within one subtype (see Table~\ref{tbl:sptind}). To summarize, we obtained spectral types of M7\,$\beta$, M9\,$\gamma$, M9\,$\gamma$, and L4\,$\gamma$ for HIP~78530~B, J1610--1913, G06214~B, and J1609--2105~b, respectively.

\begin{centering}
\begin{deluxetable*}{lcrrccc}
\tablewidth{0pt}
\tablecolumns{6}
\tablecaption{\label{tbl:sptind} Spectral type}
\tablehead{
\colhead{Object} & \colhead{SpT} & \multicolumn{4}{c}{SpT (index)} \\
\cline{3-6}
\vspace{-2mm} \\
&\colhead{Visual} &	\colhead{H$_2$O\tablenotemark{a}	}&	\colhead{H$_2$O-1\tablenotemark{b}}	&	\colhead{H$_2$O-2\tablenotemark{b}}	&	\colhead{FeH\tablenotemark{b}}	}
\startdata
HIP~78530~B	&	M7\,$\beta$	&	M7.4	$\pm$	0.5	&	M8.9	$\pm$	1.2	&	M7.5	$\pm$	0.5	&	M6.5	$\pm$	1.5	\\
J1610--1913~B	&	M9\,$\gamma$	&	M8.4	$\pm$	0.4	&	M9.5	$\pm$	1.1	&	M8.6	$\pm$	0.5	&	M8.3	$\pm$	1.5		\\
G06214~B	&	M9\,$\gamma$	&	M9.8	$\pm$	0.6	&	L0.2	$\pm$	1.2	&	M9.5	$\pm$	0.6	&	M9.8	$\pm$	1.5		\\
J1609--2105~b	&	L4\,$\gamma$	&	L2.9	$\pm$	1.0	&	L2.6	$\pm$	1.2	&	$\cdots$	&	$\cdots$				\\
\vspace{-2mm}
\enddata
\tablenotetext{a	}{From \citet{2007ApJ...657..511A}}
\tablenotetext{b	}{From \citet{2004ApJ...610.1045S}}
\end{deluxetable*}
\end{centering}

\begin{centering}																																									
\begin{deluxetable*}{llrrrr}																																									
\tablewidth{0pt}																																									
\tablecolumns{6}																																									
\tablecaption{\label{tbl:ew} Gravity scores from equivalent width}																																									
\tablehead{																																									
\colhead{Object}	&	\colhead{SpT}		&	\multicolumn{4}{c}{EW (\AA) [Gravity Score\tablenotemark{a}]}	\\																																																																					
\vspace{-2mm} \\																																									
	&	&	\colhead{\ion{Na}{1} 1.138\,$\mu$m}			&	\colhead{\ion{K}{1} 1.169\,$\mu$m}			&	\colhead{\ion{K}{1} 1.177\,$\mu$m}			&	\colhead{\ion{K}{1} 1.253\,$\mu$m}																										
\\																																									
\cline{3-6}																																								
\vspace{-2mm}																																									
}\startdata																																									
HIP~78530~B	&	M7	&	7.5	$\pm$	0.9	[	1	]	&	1.2	$\pm$	0.8	[	2	]	&	2.5	$\pm$	0.7	[	2	]	&	2.2	$\pm$	0.6	[	1	]	\\										
J1610--1913~B	&	M9	&	5.2	$\pm$	0.6	[	2	]	&	0.6	$\pm$	0.3	[	2	]	&	1.0	$\pm$	0.3	[	2	]	&	0.9	$\pm$	0.1	[	2	]	\\										
G06214~B	&	M9	&	8.7	$\pm$	1.2	[	1	]	&	2.5	$\pm$	1.5	[	2	]	&	5.1	$\pm$	1.3	[	1	]	&	1.6	$\pm$	1.2	[	2	]	\\										
J1609--2105~b	&	L4	&		\nodata		[	n	]	&		\nodata		[	n	]	&		\nodata		[	n	]	&		\nodata		[	n	]	\\							\vspace{-2mm}			
\enddata
\tablenotetext{a	}{See \citet{2013ApJ...772...79A}}
\end{deluxetable*}	
\end{centering}

\begin{centering}																																							
\begin{deluxetable*}{llrrrrcrc}																																						
\tablewidth{0pt}																																							
\tablecolumns{9}																																							
\tablecaption{\label{tbl:gravity} Gravity class}		
\tablehead{																																							
\colhead{Object}	&	\colhead{SpT}	&	\multicolumn{4}{c}{Index Values [Gravity Score\tablenotemark{a}]}															&	\multicolumn{3}{c}{Gravity}			\\	
\cline{7-9}																																							
\vspace{-2mm} \\																																							
	&		&	\colhead{FeH$_Z$}						&	\colhead{FeH$_J$}						&	\colhead{VO$_Z$}						&	\colhead{$H$-cont}						&	\colhead{Score\tablenotemark{b}}	&	\colhead{Class} &	\colhead{Visual}\\
\cline{3-4}	
\vspace{-2mm}		
}\startdata																																							
HIP~78530~B	&	M7	&	1.062	$\pm$	0.001	[	1	]	&		\nodata		[	n	]	&	1.072	$\pm$	0.002	[	n	]	&	0.987	$\pm$	0.002	[	1	]	&	1n21	&	{\sc int-g}	& $\beta$\\				
J1610--1913~B	&	M9	&	1.053	$\pm$	0.001	[	2	]	&	1.053	$\pm$	0.008	[	2	]	&	1.115	$\pm$	0.002	[	n	]	&	1.019	$\pm$	0.001	[	2	]	&	2n22	&	{\sc vl-g}	& $\gamma$\\				
G06214~B	&	M9	&	1.136	$\pm$	0.005	[	1	]	&	1.09	$\pm$	0.03	[	2	]	&	1.232	$\pm$	0.005	[	n	]	&	1.017	$\pm$	0.001	[	2	]	&	2n22	&	{\sc vl-g}	& $\gamma$\\				
J1609--2105~b	&	L4	&		\nodata		[	n	]	&	1.04	$\pm$	0.13	[	2	]	&		\nodata		[	n	]	&		\nodata		[	n	]	&	2nnn	&	{\sc vl-g}	& $\gamma$\\
\vspace{-2mm}		
\enddata																																							
\tablenotetext{a	}{See \citet{2013ApJ...772...79A}}	
\tablenotetext{b}{Respectively the scores for FeH (highest of FeH$_Z$ and FeH$_J$), VO$_Z$, alkali line (rounded mean of \ion{Na}{1} and \ion{K}{1} line scores, see Table~\ref{tbl:ew}), and $H$-cont.}					
\end{deluxetable*}																																							
\end{centering}	

We also applied the gravity classification scheme of \citet{2013ApJ...772...79A} for moderate-resolution spectra, which is based on the strength of the FeH, VO$_z$, and H$_{\rm cont}$ spectral indices and the equivalent width of \ion{Na}{1} and \ion{K}{1} lines in the $J$ band at 1.138, 1.169, 1.177 and 1.253\,$\mu$m. A score of 0 is given to objects having a value within 1$\sigma$ from the mean value of the field dwarf sequence, a score of 1 or 2 designates intermediate and very low-gravity objects, respectively, where the dividing criterion is established to roughly separate objects with optical gravity classification of $\beta$ and $\gamma$, and the score is replaced by the symbol ``n'' when the spectrum does not cover the spectral range of the index or when the index is not appropriate for the spectral type of the object. 
Table~\ref{tbl:ew} presents the equivalent widths and their respective gravity score for each alkali line. Under the scheme of \citet{2013ApJ...772...79A}, these four equivalent width scores account for one fourth of the final gravity score, as for the two FeH indices. The spectral indices and the final gravity class (field, intermediate, or very low gravity) are presented in Table~\ref{tbl:gravity}. 
For HIP~78530~B, we obtain a score of 1n21 for the FeH, VO$_z$, Alkali lines and $H$-cont indices, respectively, which classifies it as an intermediate-gravity object. J1610--1913~B and G06214~B are classified as very low-gravity objects, with both a scores of 2n22, respectively. Finally, J1609--2105~b is also classified as very low gravity, but with some reserve considering its score of 2nnn, the spectral range and low S/N preventing us from using all indices but FeH$_J$. 
The index-based gravity classes of all objects agree with our visual classification. The index-based FLD-G, INT-G, and VL-G classes defined by \citeauthor{2013ApJ...772...79A} (\citeyear{2013ApJ...772...79A}) were constructed to correspond to the $\alpha$, $\beta$ and $\gamma$ classes introduced by \citet{2005ARAA..43..195K} and \citet{2006ApJ...639.1120K}, and that we used for our visual classification. We use the latter denomination throughout this work for simplicity. 

\begin{figure}
\plotone{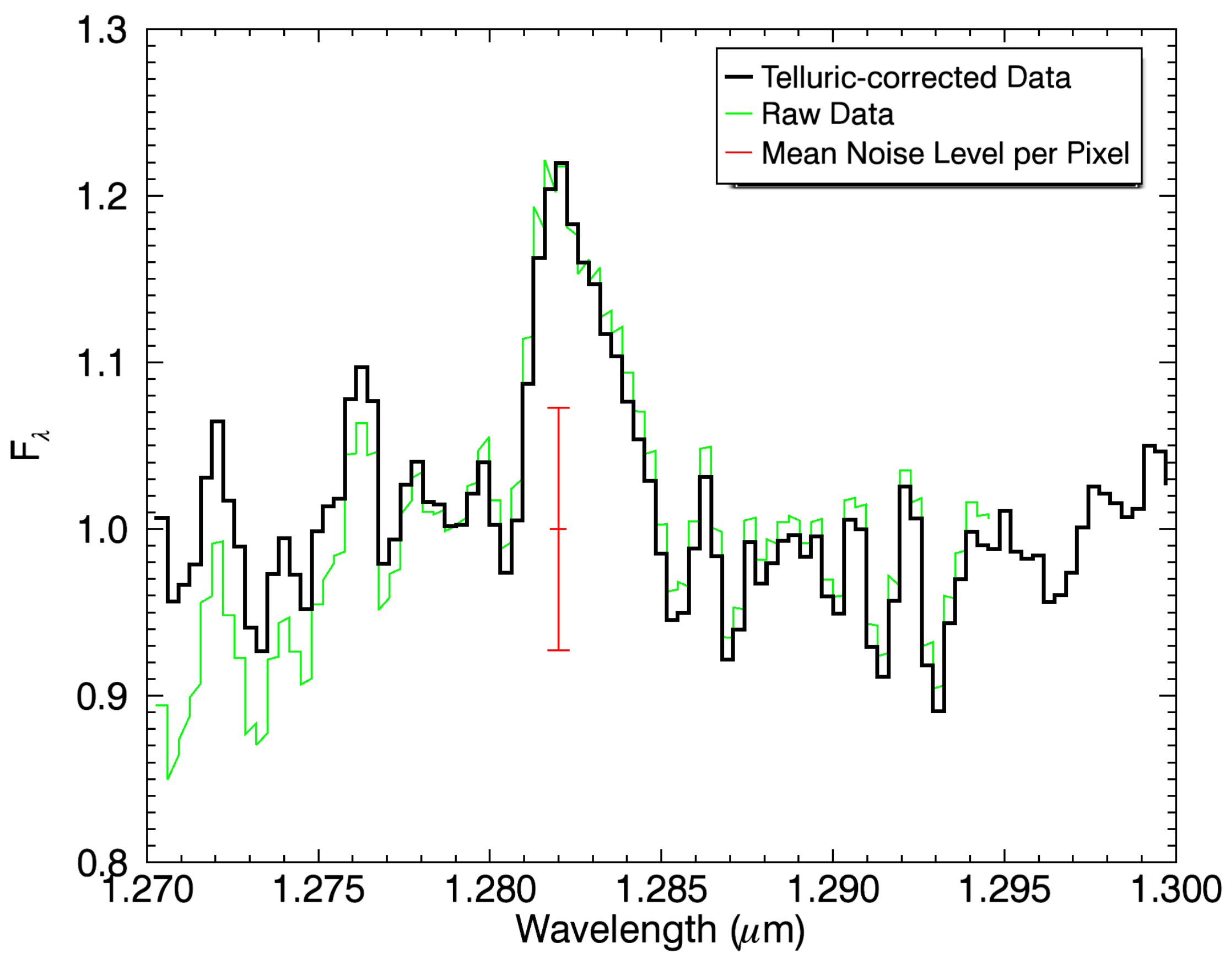}
\caption{\label{fig:pab} Zoom on the Paschen-$\beta$ line (1.282\,$\mu$m) in our spectrum of G06214~B. We measure an equivalent width of $(-4.4\pm 0.5)$\,\AA. The red error-bar represents the average noise level in the plotted region and the green spectrum is the data before telluric correction.}
\end{figure}

The Paschen-$\beta$ line at 1.282\,$\mu$m is detected in emission in the spectrum of G06214~B (see Figure~\ref{fig:pab}). This emission line was previously observed and discussed in \citet{Bowler:2011gw}. 
Bowler et al. conclude that this emission is a sign of accretion or outflow, revealing the presence of a circumplanetary disk. 
The presence of a disk is consistent with the $K-L^\prime$ excess ($1.18\pm0.10$) observed by \citet{2011ApJ...726..113I}.
Bowler et al. reported an equivalent width of $-11.4\pm0.3$\,\AA, which is significantly higher than the value that we measure here $(-4.4\pm0.5$\,\AA). This might be a sign that the accretion or outflow is variable. 
We verified that this feature is present in our raw spectrum (Figure~\ref{fig:pab}), rather that an artifact that could have been introduced by an improper correction of the Paschen-$\beta$ line in our A0 telluric standard star. 
The Brackett-$\gamma$ line at 2.166\,$\mu$m is also present in emission in our spectrum of G06214~B (EW$ = -0.24 \pm 0.05$\,\AA), providing further evidence for the presence of accretion or outflow. 

\section{COMPARISON WITH ATMOSPHERE MODELS}\label{sec:discussion} 

In the following sections, we compare the spectra of the wide companions in USco to the synthetic spectra from the {\sc BT-Settl} \citep{Allard:2011wp} and the {\sc Drift-Phoenix} \citep{2008ApJ...675L.105H,Witte:2009br,2011AA...529A..44W} models. 
Synthetic spectra with $T_{\rm eff}$ ranging from 1500\,K to 3500\,K, \logg\ ranging from 2.5 to 6.0, and solar metallicity were considered for the analysis and were binned to the same spectral resolution as our observed spectra.
A first fit was performed by minimizing the goodness-of-fit statistic ($G_k$) described in \citet{Cushing:2008kb}. 
The minimization was performed 10\,000 times, each time with a Gaussian distribution of random noise, corresponding to the uncertainties of our measured fluxes, added to our data in each resolution element. The fraction of the Monte Carlo simulations ($f_{\rm MC}$) in which the synthetic spectrum was identified as the best fitting model is then considered to evaluate the precision of the determination of \teff\ and \logg. The fit is evaluated for the whole spectrum at once, but also for each band separately.
The best fits found through this method are summarized in Table~\ref{tbl:gfit}.
In addition, the same sets of synthetic spectra were visually compared with our observations. 
The use of solar metallicity models for USco is reasonable in light of the results of \citet{2004ApJ...609..854M}, in particular see their section 4.3.3. 
We have nevertheless included BT-Settl models at higher metallicity ($[M/H]=+0.5$) and verified that our main conclusions about the companions' properties remained valid. 

\begin{centering}
\begin{deluxetable*}{cc ccc  ccc llc ccc}
\tablewidth{0pt}
\tablecolumns{14}
\tablecaption{\label{tbl:gfit} Best fit models based on the goodness-of-fit statistics}
\tablehead{
\colhead{Band} & \colhead{Model\tablenotemark{a}} & \multicolumn{3}{c}{HIP~78530~B}  & \multicolumn{3}{c}{J1610--1913~B} & \multicolumn{3}{c}{G06214~B} & \multicolumn{3}{c}{J1609--2105~b} \\
\cmidrule(r){3-5}
\cmidrule(r){6-8}
\cmidrule(r){9-11}
\cmidrule(r){12-14}
\colhead{} & \colhead{} & \colhead{\teff} & \colhead{\logg} & \colhead{$f_{\rm MC}$} & \colhead{\teff} & \colhead{\logg} & \colhead{$f_{\rm MC}$} & \colhead{\teff} & \colhead{\logg} & \colhead{$f_{\rm MC}$} & \colhead{\teff} & \colhead{\logg} & \colhead{$f_{\rm MC}$} }
\startdata
$J$ & {\sc BT-S}  & 2800 & 3.5 & 0.93 & 2600 & 3.5 & 1.00 & 2500\tablenotemark{b} & 3.5\tablenotemark{b} & 0.94 & 1600 & 3.5 & 1.00 \\
$J$ & {\sc D-P}  & 2900 & 5.0 & 1.00 & 2700 & 4.5 & 1.00 & 2500 & 5.0 & 0.83 & 1600 & 3.0 & 1.00 \\[2mm]

$H$ & {\sc BT-S}  & 2900 & 4.5 & 0.61 & 2700 & 3.5 & 1.00 & 2600 & 3.5 & 0.50 & 1600 & 4.0 & 1.00 \\
$H$ & {\sc D-P}  & 2900 & 4.5 & 1.00 & 2600 & 3.5 & 1.00 & 2600\tablenotemark{c} & 3.0\tablenotemark{c} & 0.69 & 1800 & 3.0 & 1.00 \\[2mm]

$K$ & {\sc BT-S}  & 2600 & 2.5 & 0.88 & 2700 & 3.5 & 1.00 & 2600 & 2.5 & 0.94 & 1700 & 4.0 & 1.00 \\
$K$ & {\sc D-P}  & 2800 & 4.5 & 0.67 & 2600 & 4.0 & 0.78 & 2300 & 3.0 & 1.00 & 1800 & 3.5 & 1.00 \\[2mm]

$JHK$ & {\sc BT-S}  & 2600 & 3.0 & 1.00 & 2400 & 3.0 & 1.00 & 2100\tablenotemark{b} & 3.0 & 1.00 & 1600 & 3.5 & 1.00 \\
$JHK$ & {\sc D-P}  & 2600 & 3.5 & 1.00 & 2300 & 3.0 & 1.00 & 2100 & 3.0 & 1.00 & 1600 & 3.0 & 1.00 \\ 
\vspace{-2mm}
\enddata
\tablenotetext{a}{{\sc BT-S}: {\sc BT-Settl}, {\sc D-P}:{\sc Drift-Phoenix}}
\tablenotetext{b}{The VO band was omitted from the fit.}
\tablenotetext{c}{There is also a local minimum at 1700\,K but this model is clearly not appropriate for other bands, we have thus restricted the range of temperatures for the fit to $>1800$\,K.}
\end{deluxetable*}
\end{centering}

\begin{figure*}
\plotone{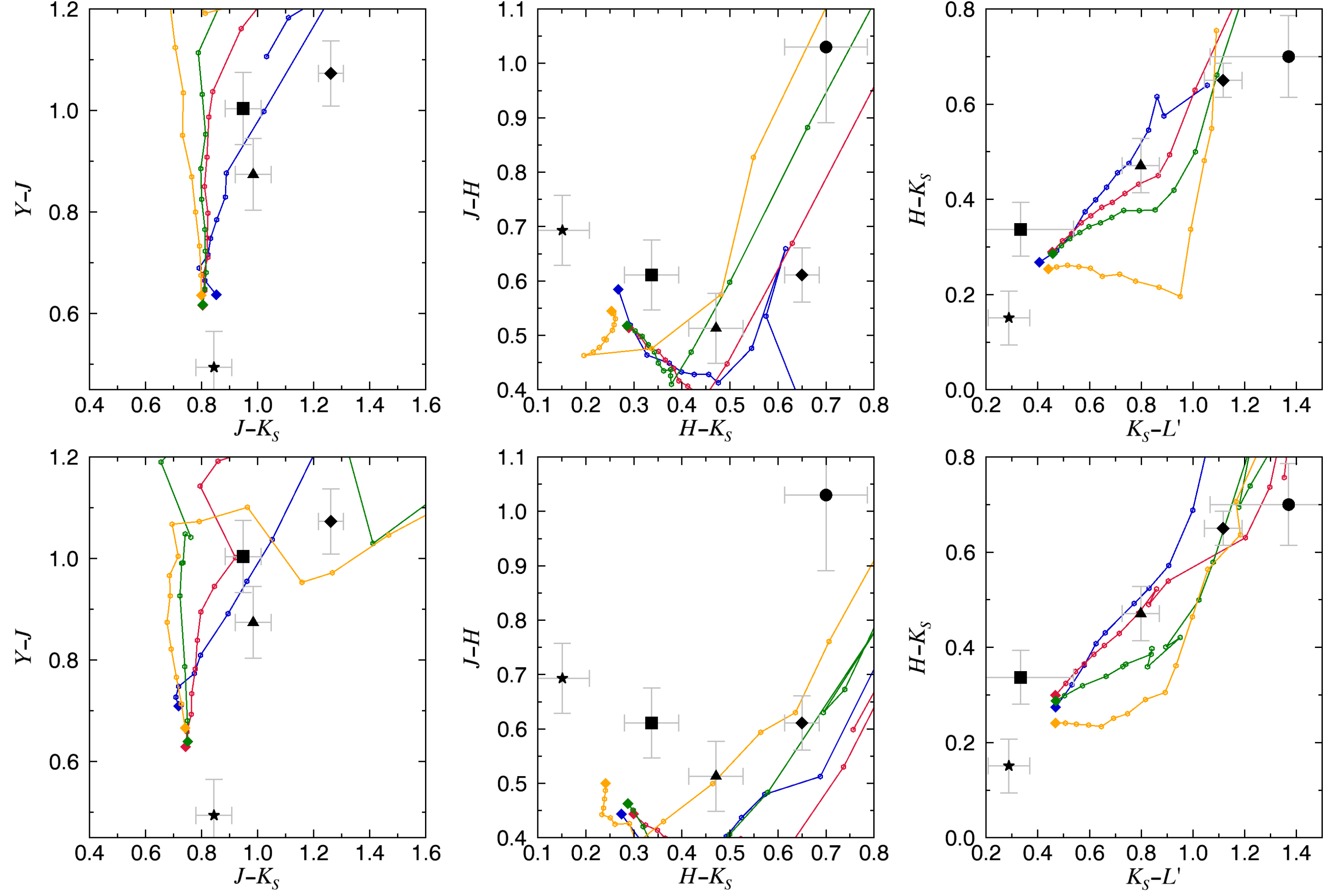}
\caption{\label{fig:colcol} 
Colors of HIP~78530~B (square), J1610--1913~Ab and B (star and triangle, respectively), G06214~B (diamond) from this work, and of J1609--2105~b from \citet{2010ApJ...719..497L} (circle). The colors of HIP~78530~B and J1610--1913~Ab and B have been corrected for extinction (see text). The solid lines on the top row show synthetic colors from the {\sc BT-Settl} models for \logg\ of 3.0 (blue), 4.0 (red), 4.5 (green), and 5.5 (orange), and for temperatures ranging from 3000\,K to 1600\,K by 100\,K increments. The solid lines on the bottom row shows synthetic colors from the {\sc Drift-Phoenix} models for \logg\ of 3.0 (blue), 4.0 (red), 5.0 (green), and 6.0 (orange), for the same temperatures. 
}
\end{figure*}

We also compare our photometric measurements to synthetic magnitudes calculated from the two sets of synthetic spectra.
To compute the synthetic magnitudes, we used the filter profiles\footnote{http://irtfweb.ifa.hawaii.edu/{\textasciitilde}nsfcam/filters.html} and 
the magnitude zero points from \citet{2005PASP..117.1459T} and online supplements\footnote{http://irtfweb.ifa.hawaii.edu/IRrefdata/iwafdv.html}.
We compared the synthetic and observed magnitudes and determined the best-fit model by minimizing the $\chi ^2$ over the $YJHK_{\rm s}L^\prime$ bands. In Figure~\ref{fig:colcol}, we also compare the observed and model colors. As visible on the figure, the measured colors of the companions agree reasonably well with the colors expected from the models. In particular, the companions colors seem to roughly reproduce the shapes of the model curves. The only noteworthy discrepancy is a systematic offset of up to $\sim$\,0.1\,mag in $J-H$. The relative positions of the colors of the companions in the different color-color diagrams, when compared to the model curves, readily indicate the relative temperatures of the companions. From the two rightmost columns of the figure, we get, respectively from the hottest to the coldest, HIP~78530~B, J1610--1913~B, G06214~B and J1609--2105~b. This ordering is consistent with the spectral types presented above. In almost all panels of Figure~\ref{fig:colcol}, we can also see that the colors of the companions are closer to the models of low surface gravity, as expected for young objects. The only panel where this is not the case is $J-H$ vs $H-K_{\rm s}$, although it seems that this problem would disappear if the $\sim$\,0.1\,mag systematic offset in $J-H$ mentioned earlier could be resolved. 

The temperature estimates based on all of these analyses, along with comments on the agreement with the models, are discussed in Sections~\ref{sec:D78530}, \ref{sec:D1610}, \ref{sec:D06214}, and \ref{sec:D1609} for each object separately.

\subsection{HIP~78530 B}\label{sec:D78530}

\begin{figure*}
\includegraphics[width=\textwidth]{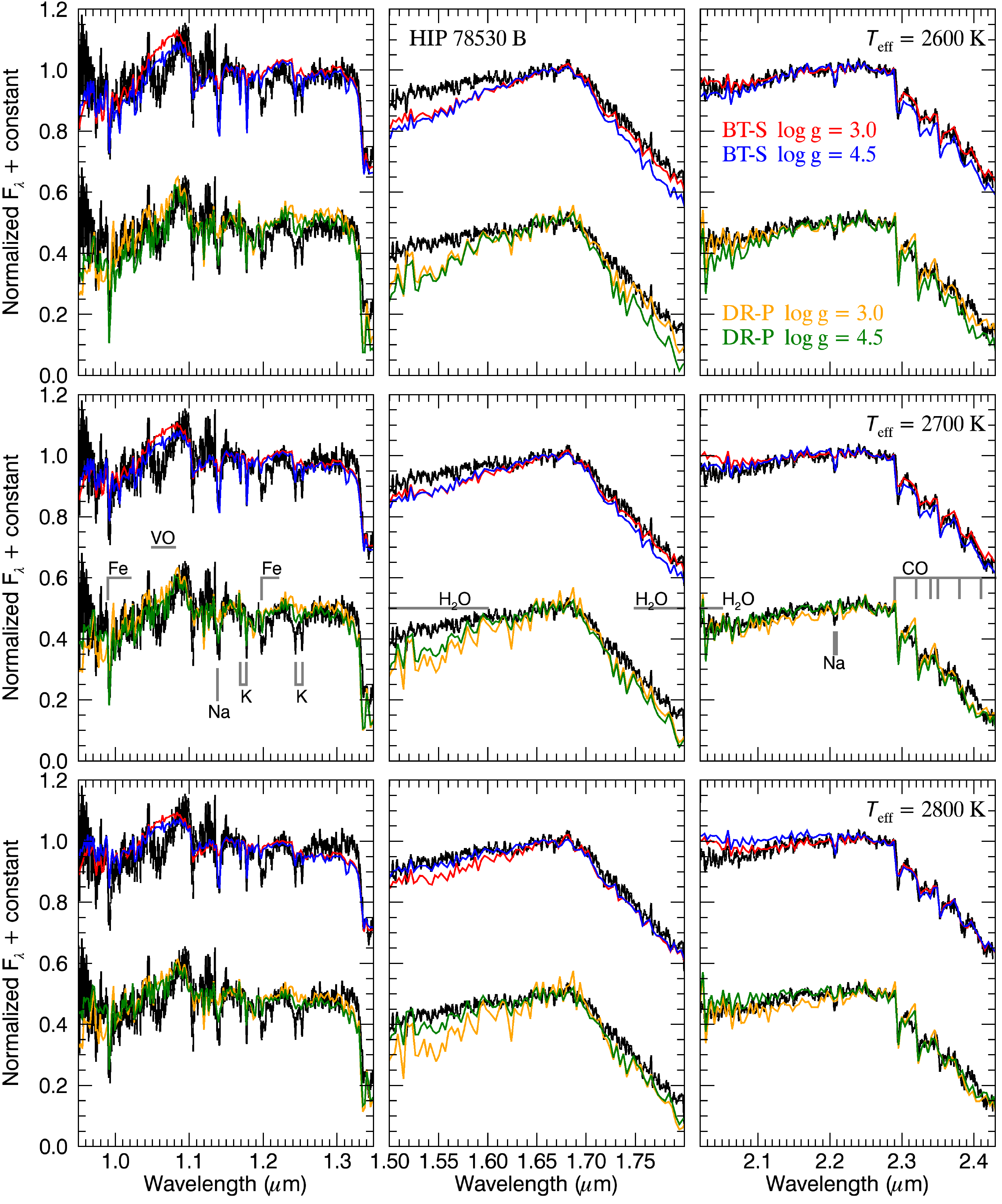}
\caption{\label{fig:78530mod} GNIRS spectrum of HIP~78530~B (black) corrected for an extinction of $A_V=0.5$ and compared with synthetic spectra of {\sc BT-Settl} at $\logg$ of 3.0 (red) and 4.5 (blue) and {\sc Drift-Phoenix} at $\logg$ of 3.0 (yellow) and 4.5 (green) for $\teff$=2600\,K (top row), $\teff$=2700\,K (middle row), and $\teff$=2800\,K (bottom row).}
\end{figure*}

Figure~\ref{fig:78530mod} compares the observed spectrum of HIP~78530~B to synthetic spectra selected from the grid of {\sc BT-Settl} and {\sc Drift-Phoenix} models. 
For the {\sc BT-Settl} models, the shape of both the $J$ and $K$ bands are better fitted by the 2600--2700\,K models, at a low gravity ($\logg = 3.0$) to match the CO lines depths, but the $H$ band is then too triangular. An effective temperature of 2800\,K is needed to get the right slopes in $H$ band. The {\sc Drift-Phoenix} models at 2600\,K and low gravity are able to better reproduce the features in the $J$ band, particularly the VO band at 1.06\,$\mu$m. At higher temperatures (2800\,K), this feature is not deep enough in the models. The $K$ band is well fitted in the 2600--2800\,K temperature range, with a marginally better fit at 2600\,K.
For the $H$ band, a temperature even higher than 2800\,K is needed to correctly fit the shape of the pseudo-continuum with the {\sc Drift-Phoenix} models.
Using the goodness-of-fit statistics (Table~\ref{tbl:gfit}), the best fits for the individual bands are for 2600--2900\,K, in good agreement with the above estimates, while it is 2600\,K for the fit to the entire $JHK$ spectrum. 
As for the broadband photometry only, the best fit is achieved with $\teff$ of 2300--2700\,K and $\logg=3.5$.
Considering all of these elements, we assign a temperature of $2700\pm100$\,K to HIP~78530~B. 

The best fits discussed above occur for models at low surface gravity, in agreement with the young age of the region and with the values of gravity-sensitive spectral indices found earlier. In particular, the spectral indices for the FeH molecular bands at 0.998\,$\mu$m and 1.200\,$\mu$m are significantly weaker than those of field dwarfs, a sign of low surface gravity. 
As mentioned previously, the VO band at 1.06\,$\mu$m is gravity-sensitive. Systematically, the {\sc Drift-Phoenix} models provide a much better fit of this feature than the {\sc BT-Settl} model, although they do not significantly discriminate the surface gravity parameter. The depth of the CO molecular bands in the red part of the $K$ band does require a low gravity to be well fitted. The gravity-sensitive \ion{Na}{1} doublet at 2.206 and 2.209\,\AA\ is clearly visible in the data, but a \logg\ of 4.5 is not high enough for the models to reproduce its depth. The models are thus under predicting the depth on the Na feature.

Based on the photometry of HIP~78530~B, \citet{2013ApJ...767...31B} estimated a temperature of 3300K--3400\,K and a spectral type of $\sim$\,M3, raising the possibility that the companion was instead a background star. While this higher temperature estimate could be consistent with the $K-L^\prime$ color of the companion, according to models, it would not be appropriate for the other colors (see Figure~\ref{fig:colcol}). Also, this estimate is inconsistent with our observed spectrum and the spectrum from \citet{2011ApJ...730...42L}. In addition, our observations show signs of low gravity, common proper motion (see \ref{sect:phot-astrom}), and a spectral type of M7. For all of these reasons, we rule out the possibility that this is a background star.

\subsection{[PGZ2001]~J161031.9-191305 B and Ab}\label{sec:D1610}

\begin{figure*}
\includegraphics[width=\textwidth]{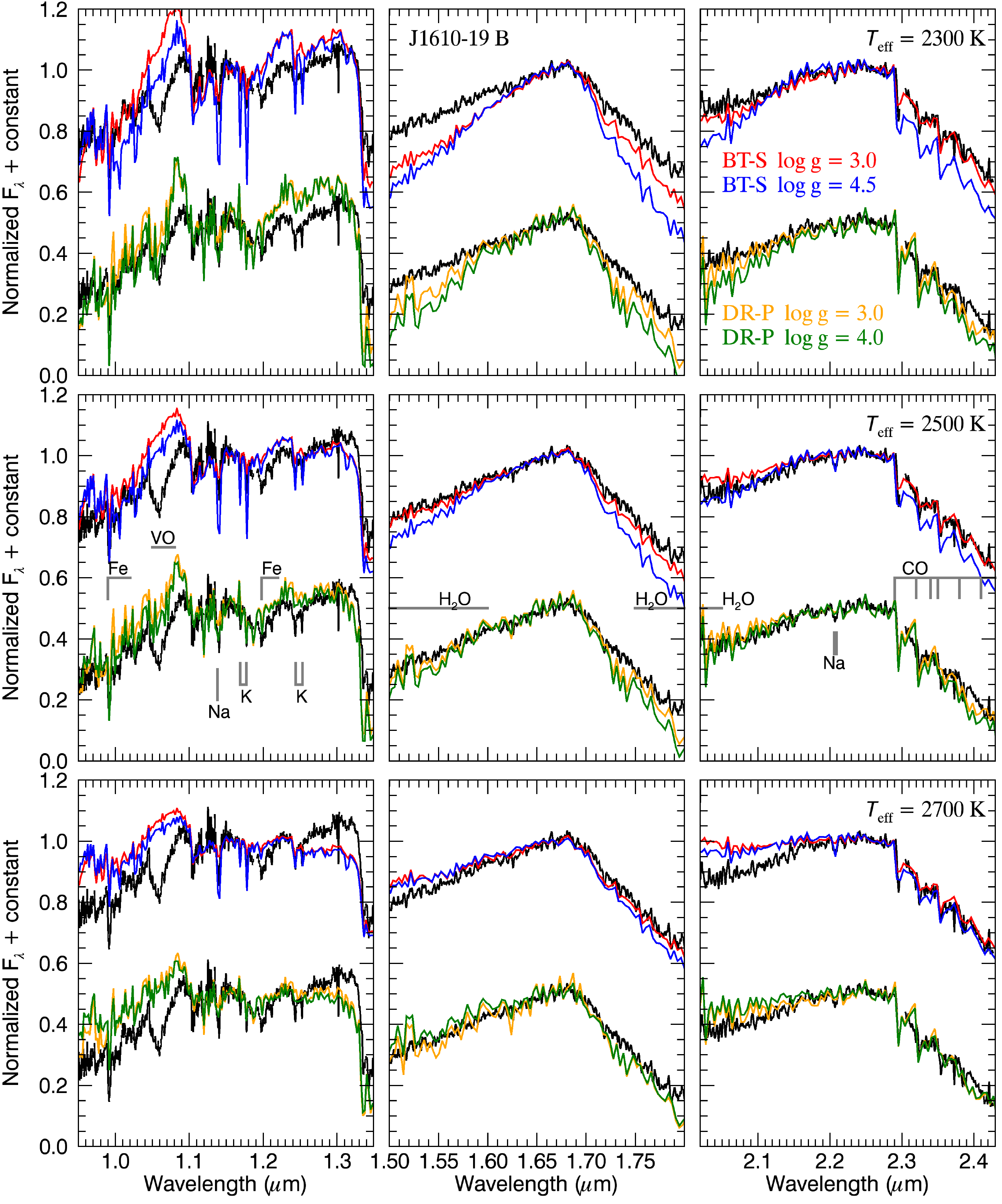}
\caption{\label{fig:1610mod} GNIRS spectrum of J1610--1913~B (black) corrected for an extinction of $A_V=1.1$ and compared with {\sc BT-Settl} synthetic spectra for $\logg$ of 3.0 (red) and 4.5 (blue) and {\sc Drift-Phoenix} synthetic spectra for $\logg$ of 3.0 (yellow) and 4.0 (green), for $\teff$=2300\,K (top row), $\teff$=2500\,K (middle row), and $\teff$=2700\,K (bottom row).}
\end{figure*}

A comparison of the spectrum of J1610--1913~B with various model spectra is shown in Figure~\ref{fig:1610mod}.
The $K$ band is best fitted by the {\sc BT-Settl} model at 2300\,K, at low surface gravity ($\logg = 3.0$), and by the {\sc Drift-Phoenix} model at 2500\,K, for either surface gravities shown. Hotter models (2700\,K) fail to match the blue side of this band.
In the $H$ band, the best fit occurs for temperatures of 2500--2700\,K, with a notably better fit at low surface gravity for 2700\,K.
In the $J$ band, the best fit of the pseudo-continuum as well as the depth of the water band at 1.33\,$\mu$m seems to take place at 2500\,K, for either surface gravities shown, although the VO band at 1.06\,$\mu$m is not quite deep enough at this temperature in the models. The depth of the VO feature is best matched by the 2300\,K {\sc Drift-Phoenix} models, for either surface gravities, but then the fit is not as good in the other parts of the $J$ band. Again, the {\sc BT-Settl} models fail to reproduce the VO band at any temperature or gravity. 
The best fits for the individual bands using the goodness-of-fit method (Table~\ref{tbl:gfit}) are found for temperatures of 2600--2700\,K, and the fit to the entire $JHK$ spectrum indicates a temperature of 2300--2400\,K. All of these best fits occur for a  $\logg$ equal to or less than 4.5.
For the fit of the broadband magnitudes, after proper correction for extinction as mentioned previously, we get a temperature of 2300\,K with a $\logg$ of 3.0 for both models; this is close to the simultaneous $JHK$ model fit result. 
Considering all of these values, we assign an effective temperature of $2500\pm200$\,K to J1610--1913~B. The spectral and the broadband photometry fits both favour a very low surface gravity for this object, in good agreement with the values of the FeH index and alkali lines equivalent widths calculated earlier.

We have not observed the closer-in companion (Ab) in the J1610--1913 system using spectroscopy, as getting a contamination-free spectrum of this object with a source $\sim$\,3 mag brighter at a separation of only $\sim$\,0.2\,\arcsec\ is too challenging for the instrumental setup we used. Nevertheless, we have photometric measurements from our imaging and we can compare those with the models to assess its effective temperature. With the photometric points obtained, the best fit to the synthetic magnitudes would indicate an effective temperature of $2900\--3300$\,K with a $\logg$ of 3.5.

\subsection{GSC 06214-00210 B}\label{sec:D06214}

\begin{figure*}
\includegraphics[width=\textwidth]{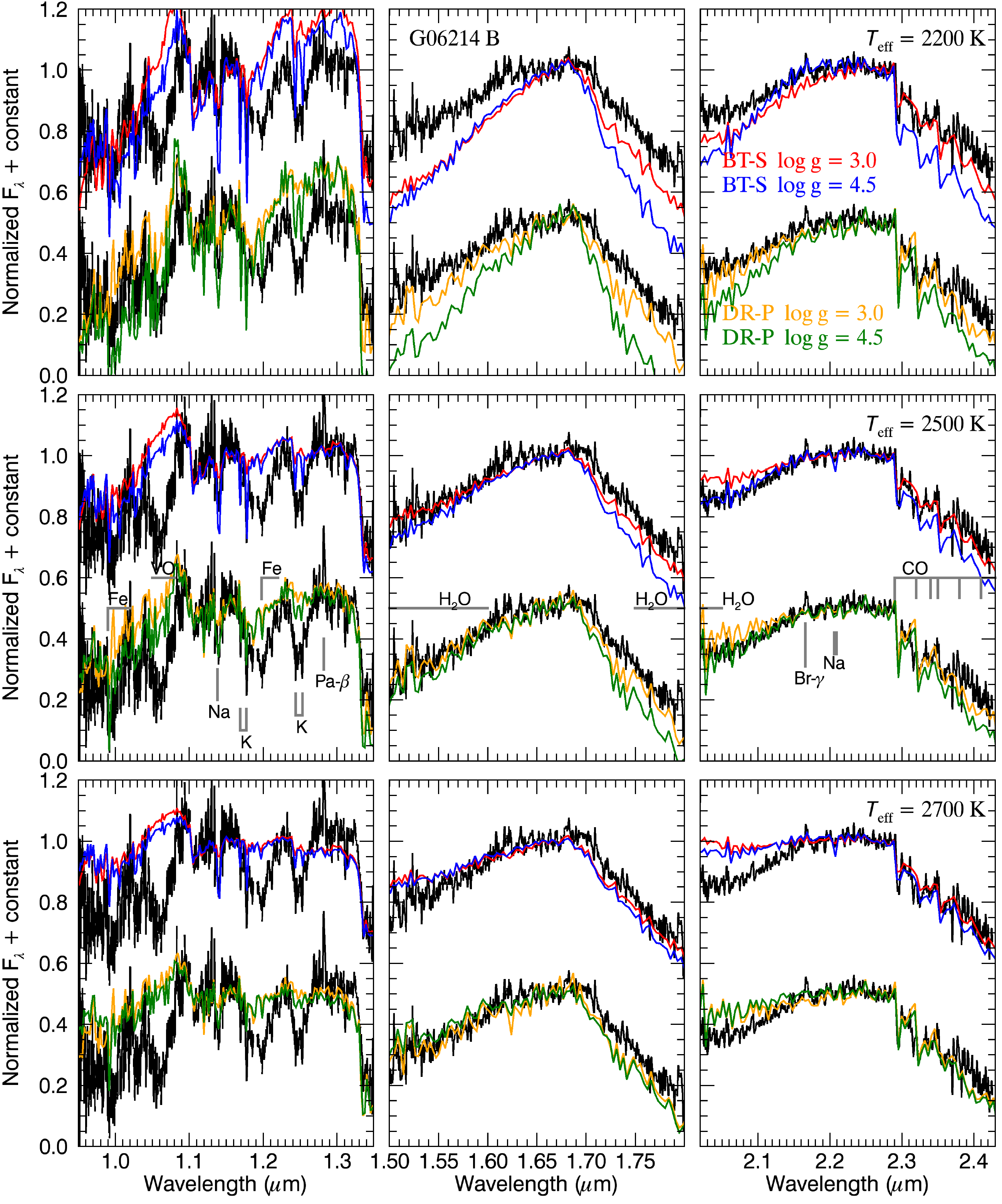}
\caption{\label{fig:06214mod} GNIRS spectrum of G06214~B (black) compared with {\sc BT-Settl} synthetic spectra for $\logg$ of 3.0 (red) and 4.5 (blue) and {\sc Drift-Phoenix} synthetic spectra for $\logg$ of 3.0 (yellow) and 4.5 (green), for $\teff$=2200\,K (top row), $\teff$=2500\,K (middle row), $\teff$=2700\,K (bottom row).}
\end{figure*}

Figure~\ref{fig:06214mod} shows our GNIRS spectrum of G06214~B compared with a selection of models from the {\sc BT-Settl} and {\sc Drift-Phoenix} models. 
The $K$ band is best reproduced by models at 2200\,K and low \logg\ for both {\sc BT-Settl} and {\sc Drift-Phoenix}.
For the $H$ bands, models of 2500--2700\,K provide reasonable fits, with little effects from surface gravity.
In the $J$ band, the VO and FeH (1.2\,$\mu$m) bands are better matched by the models at 2200\,K, with {\sc Drift-Phoenix} providing a much better fit than {\sc BT-Settl}. The most important difference is the deeper water absorption band at 1.33\,$\mu$m for the models, as compared with the observations.
The goodness-of-fit evaluation (Table~\ref{tbl:gfit}) indicates temperatures of 2300--2600\,K when applied to individual bands, and 2100\,K when applied globally.
The colors of this companion and the corresponding photometric magnitudes are in excellent agreement with the models for a temperature of 2200\,K and very low \logg. 
We assign a temperature of $2300\pm200$\,K to G06214~B.

Visually, the models with $\logg$ of 3.0 or less are in better agreement with the observed spectrum, especially in the $K$ band. The same result is obtained for the fit 
of the spectrum and the broad band fluxes. 
Also for this object, the depth of the gravity-sensitive \ion{K}{1} doublet at 1.244 and 1.252\,$\mu$m in the models is insufficient to agree with the observations, even if properly degraded to the resolution of the observations. 

\subsection{1RXS~J160929.1--210524~b}\label{sec:D1609}

\begin{figure*}
\includegraphics[width=\textwidth]{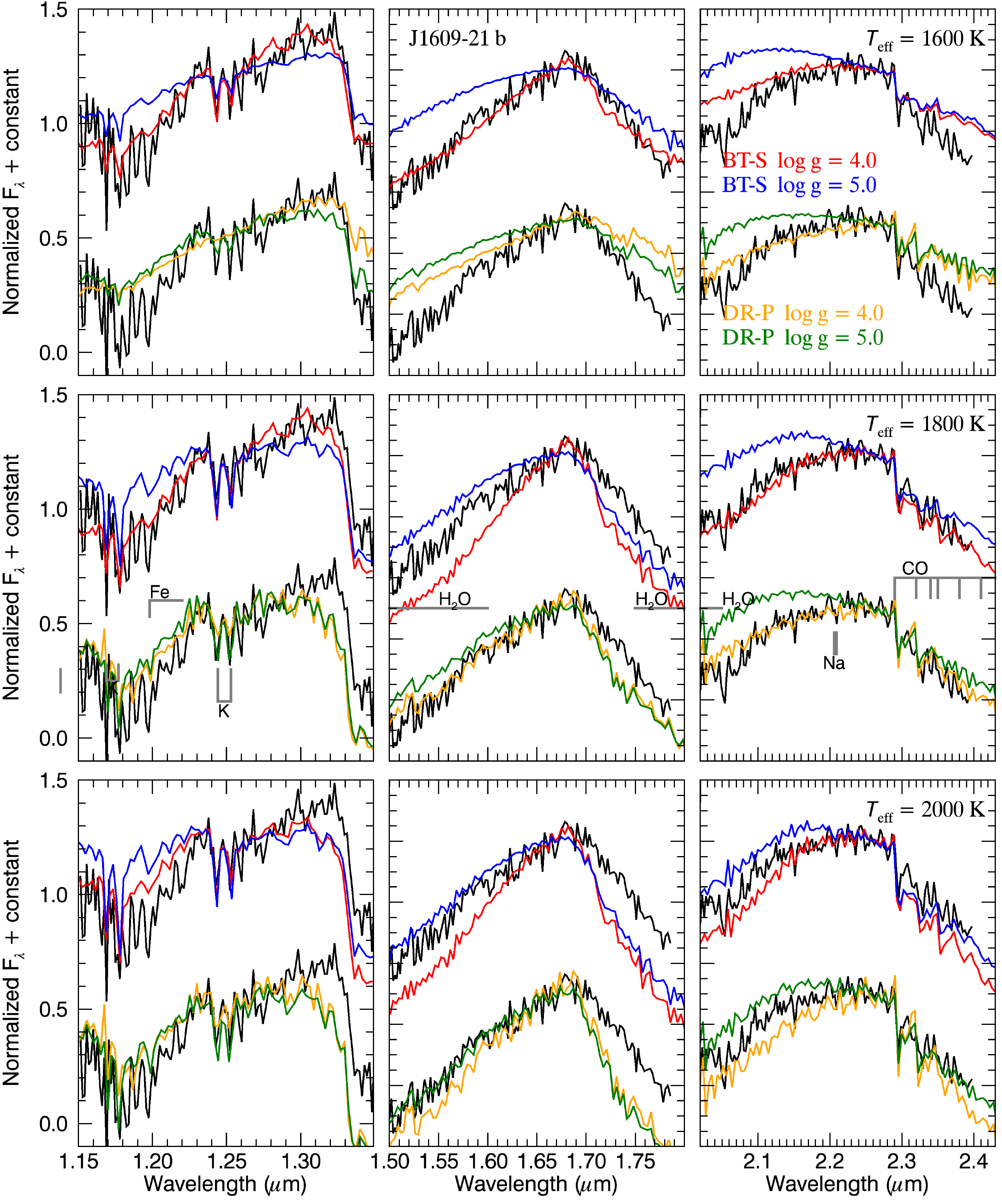}
\caption{\label{fig:1609mod} Spectrum of J1609--2105~b as observed with NIFS in the $J$ band \citep{2010ApJ...719..497L} and NIRI in the $H$ and $K$ bands \citep{2008ApJ...689L.153L}.
The observed spectrum is compared with synthetic spectra from the {\sc BT-Settl} models at $\logg$ of 4.0 (red) and 5.0 (blue) and the {\sc Drift-Phoenix} models at $\logg$ of 4.0 (yellow) and 5.0 (green), for $\teff$=1600\,K (top row), $\teff$=1800\,K (middle row), $\teff$=2000\,K (bottom row).}
\end{figure*}

Figure~\ref{fig:1609mod} presents the NIFS $J$ band \citep{2010ApJ...719..497L} and NIRI $H$ and $K$ bands \citep{2008ApJ...689L.153L} spectrum of J1609--2105~b in comparison to synthetic spectra with \teff\ ranging from 1600\,K to 2000\,K, from both the {\sc BT-Settl} and the {\sc Drift-Phoenix} models. 
The lower-gravity ($\logg=4.0$) {\sc Drift-Phoenix} model at 1800\,K gives the overall best fit, although the red side of the $J$ band is a bit too low and the slope on the red side of the $H$ band is a bit too steep. 
The 1800\,K {\sc BT-Settl} model does not provide as good a fit as the 1800\,K {\sc Drift-Phoenix} model, particularly in the $H$ band where it is too peaked compared with the observed spectrum.
At lower temperatures ($\teff$=1600\,K), both models clearly fail to reproduce the observations in all bands. 
At higher temperatures ($\teff$=1800\,K), the fits are not too bad for both models, although the water absorption band in $J$ is too strong in the models and the red side of the $H$ band is too steep.
The goodness-of-fit evaluations (Table~\ref{tbl:gfit}), both band-by-band and over the whole spectrum, generally agree on a temperature of 1600--1800\,K and a \logg\ of $3.0 - 4.0$. Only the $H$ and $K$ bands fit a higher temperature of 1800\,K with the {\sc Drift-Phoenix} models. 
As for the broad band magnitudes, they yield a best-fit temperature of 1700\,K for {\sc BT-Settl} and 1800\,K for {\sc Drift-Phoenix}, in both cases with a \logg\ of 3.0. We thus assign a temperature of $1700\pm100$\,K to J1609--2105~b. The spectral indices calculated earlier classified this object as having a very low gravity; this is in good agreement with the best fits with the models obtained here. However here again, the gravity-sensitive potassium lines in $J$ are not deep enough for the {\sc Drift-Phoenix} model at lower gravity.

\subsection{Mass estimates}\label{ssec:mass}

\begin{figure}
\includegraphics[width=\linewidth]{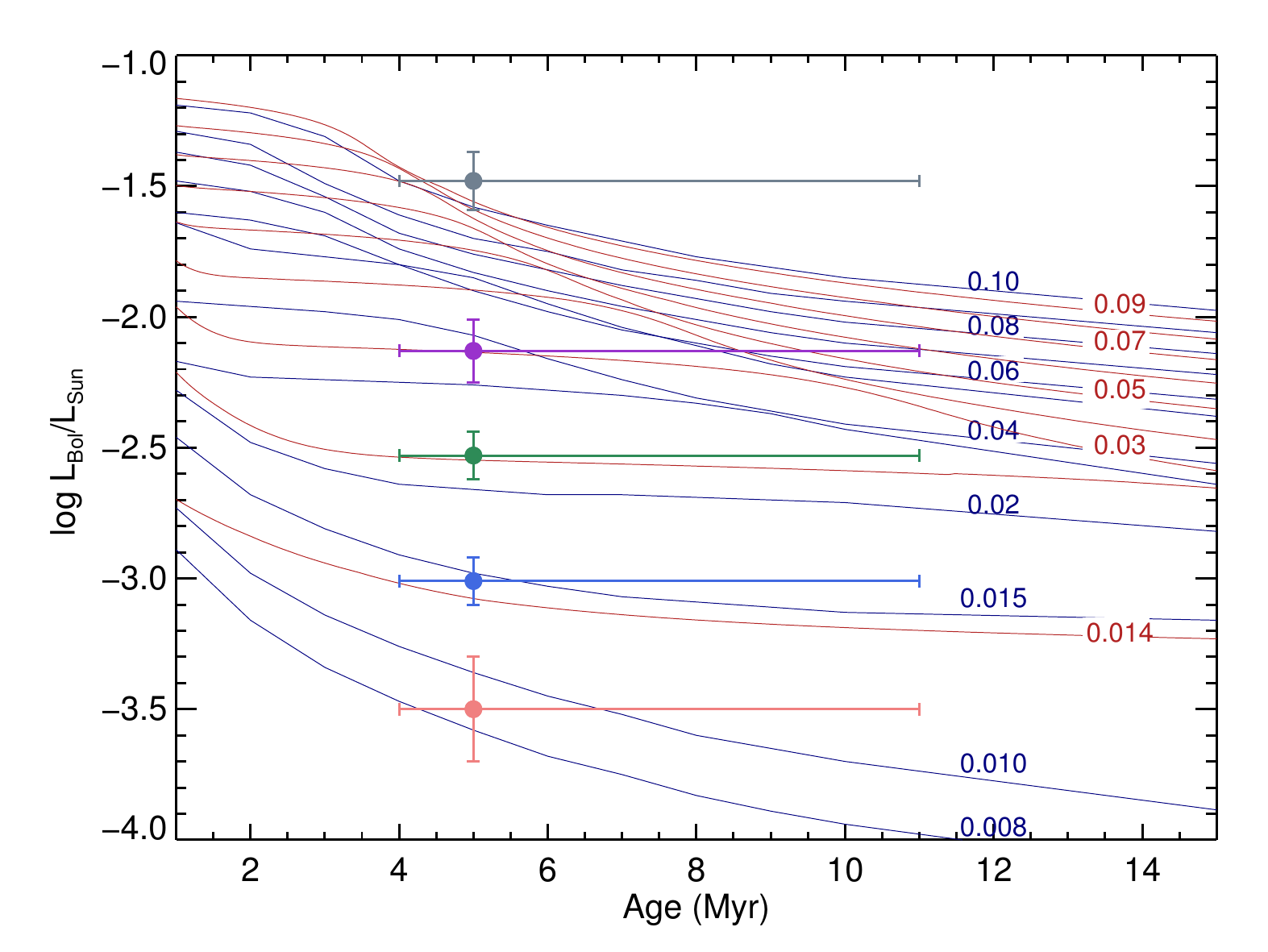}
\caption{\label{fig:lbol} Luminosity for different masses (labeled in units of \msun) as a function of age from the evolution models of \citet{1997ApJ...491..856B} (red) and \citet{Allard:2013uv} (blue). The points with error bars mark the estimated luminosities of J1609--2105~b (pink), G06214~B (cyan), HIP~78530~B (green), J1610--1913~B (purple), and J1610--1913~Ab (grey).}
\end{figure}

We have estimated the masses of the companions using two different approaches, comparing either their estimated bolometric luminosities or their estimated effective temperatures to the predictions of evolution models. We have used two sets of evolution models, the models from  \citet{1997ApJ...491..856B}  and the models from \citet{Allard:2013uv}, the latter being based on the CIFIST2011 {\sc BT-Settl} atmosphere models and the AMES-Cond isochrones \citep{Baraffe:2003bj}.
For the effective temperatures, we simply used the estimates presented in the previous section. 
For each object, a synthetic bolometric luminosity was computed for every synthetic spectrum within the range of plausible temperature and gravity determined previously. 
This was done by first scaling the model spectrum to the observed flux level, by minimizing the $\chi^2$ between the synthetic magnitudes of the model spectrum and the ones we observed, and then by integrating the entire model spectrum and converting the total flux to luminosity using the mean USco distance of 145\,pc \citep{1999AJ....117..354D} and 14\,pc uncertainty (as discussed in \citealt{2011ApJ...726..113I}). 
The error on luminosity is chosen to be large enough to include all the temperatures and gravity ranges described above and encompass results from both atmosphere models.
The resulting luminosities are included in Tables~\ref{tbl:78530}, \ref{tbl:1610}, \ref{tbl:06214}, and \ref{tbl:1609}; see also Figure~\ref{fig:lbol} for a comparison of these luminosities with the prediction of evolution models.

Estimating the masses based on the calculated luminosities and an age of 5\,Myr, we obtain
$0.008\pm0.001$\,\msun\ for J1609--2105~b, 
$0.015\pm0.001$\,\msun\ for G06214~B, 
$0.032\pm0.004$\,\msun\ for J1610--1913~B, 
$0.022\pm0.001$\,\msun\ for HIP~78530~B, and 
$0.12\pm0.02$\,\msun\ for J1610~Ab. 
For an age of 10\,Myr, the corresponding masses are respectively, 
$0.011\pm0.001$\,\msun , 
$0.016\pm0.001$\,\msun , 
$0.058\pm0.011$\,\msun , 
$0.023\pm0.002$\,\msun , and 
$0.16\pm0.02$\,\msun . 
The impact of the revised older age proposed by \cite{Pecaut:2012ux} is relatively small for the three lowest mass objects, G06214~B and HIP78530~B being in a relatively stable deuterium burning phase.
The impact for J1610--1913~Ab and B would be more important. 
 
With the objects ordered according to increasing effective temperature, as above, it is obvious that something is off for J1610--1913~B. Namely, its luminosity is much higher than expected. Indeed, J1610--1913~B has an estimated temperature of 2500\,K, cooler than HIP~78530~B at 2700\,K, but its luminosity ($\log{(L/L_\odot)} =-2.13$) comes out significantly brighter than that of HIP~78530~B ($\log{(L/L_\odot)} =-2.53$). A possible reason for this discrepancy is that the true (unknown) distance of J1610--1913~B differs largely from the mean distance of USco members. If its true distance were toward the closer side of the association, at $\sim$\,115\,pc, rather than the assumed distance of $145$\,pc, then its luminosity would be closer to $\log{(L/L_\odot)} =-2.45$. Another possibility is that the wide companion is itself an unresolved equal-mass binary. That would bring its intrinsic luminosity down by a factor of 2, to $\log{(L/L_\sun)} = -2.5$. Both effects combined would be more than enough to solve the problem. Other possibilities could include an unusually large radius, for example.
\citet{2013ApJ...773...63A} also observed and reported the J1610--1913~B over-luminosity problem, and concluded that J1610--1913~B does not look coeval with its host star. They also concluded the same for the five companions they observed in USco.
In our study, however, we observed an over-luminosity only for J1610--1913~B, the other companions luminosities being consistent with the 5--10\,Myr isochrones in a luminosity--effective temperature diagram.
If the luminosity problem mentioned above for this companion were to be resolved, the difference between its mass determined at 5 and 10\,Myr would be largely reduced.

On the other hand, the masses can be estimated directly from the evolution models by using the temperatures evaluated from the model atmosphere fits presented above.
Specifically, we can find the masses for which the evolution models predict these temperatures. 
With this method, for an age of 5\,Myr we obtain
$0.007\pm0.001$\,\msun\ for J1609--2105~b, 
$0.015\pm0.003$\,\msun\ for G06214~B, 
$0.020\pm0.006$\,\msun\ for J1610--1913~B, 
$0.029\pm0.012$\,\msun\ for HIP~78530~B, and
$0.15^{+0.31}_{-0.11}$\,\msun\ for J1610~Ab. 
For an age of 10\,Myr, the corresponding masses are respectively, 
$0.010\pm0.001$\,\msun , 
$0.016\pm0.003$\,\msun , 
$0.020\pm0.005$\,\msun , 
$0.030\pm0.013$\,\msun , and
$0.14^{+0.31}_{-0.10}$\,\msun . 
The errors encompass temperatures estimates from both sets of atmosphere models considered here. 
The masses of J1609--2105~b and G06214~B are approximately the same when evaluated from luminosity or temperature, and they agree well with the estimates made by \citet{2010ApJ...719..497L} and \citet{2011ApJ...726..113I} respectively.
The mass of J1610--1913~B is significantly lower when estimated using only its temperature, as was expected from the above comments. The difference between the two estimates exceeds the quoted uncertainties, probably indicating that there is a real problem with the brightness of this object.
For HIP~78530~B, this result is significantly higher than the previous value, but it comes with a large uncertainty and the two values can be reconciled.
The mass of the close binary J1610--1913~Ab seems to be in the stellar regime using both methods.

\section{DISCUSSION AND CONCLUDING REMARKS}

Our homogeneous comparison between the observed spectra of young substellar companions in USco and synthetic spectra from the {\sc BT-Settl} and {\sc Drift-Phoenix} models has revealed some interesting and systematic trends. 
First, the models do not succeed in reproducing the details of the spectra across the 1--2.4\,$\mu$m range simultaneously. 
As noted by \cite{Cushing:2008kb} in the case of early L-type dwarfs in the field, the best fit in the individual bands typically occur for models of different temperatures. 
At the temperatures providing the best fits in the $J$ and $K$ bands, the synthetic spectra have significantly steeper slopes in $H$ than the observed spectra, both at the blue and red ends.
An even more evident feature that is not reproduced by the models is the VO band in $J$. 
The VO absorption band at 1.06\,$\mu$m is only reproduced by the {\sc Drift-Phoenix}s models, and generally only at a temperature slightly lower than the temperature leading to the best fit in other parts of the $J$ band. The {\sc BT-Settl} models simply fail to reproduce this VO feature at any reasonable range of \teff\ and \logg .
Similar conclusions about the VO feature and the fit of the $H$ band for the {\sc BT-Settl} models were reported by \citet{Allers:2013vv}. The alkali \ion{K}{1} and \ion{Na}{1} lines in the observed spectra are systematically stronger than in the models. Note however that we have not carefully investigated the effect of metallicity on these features. 
The excellent agreement of the spectra of free floating BD in Upper Scorpius with our spectra (see Figure~\ref{fig:compare}) provides yet another argument that the above trends are common features of young BDs and point to a real shortcoming of the models. 

Also, the best temperature estimates obtained by matching the broadband magnitudes and colors of the objects to synthetic magnitudes from the models are systematically lower, by $\sim$\,200\,K on average, than the temperatures obtained from band-by-band comparisons of the spectra with models. This is also the case for the spectral fit applied globally (simultaneously across the near-infrared range), as it too is affected by the broad band colors. 

Beyond their use for testing atmosphere and evolution models, the wide low-mass substellar companions studied here are of high interest for the study of planet and star formation. In principle, such low mass companions could form like stars, through the collapse and fragmentation of a pre-stellar core, or as planets within a circumstellar disk, but their combination of low mass and wide separation poses a challenge to both processes. A formation in-situ within the circumstellar disk of the primary would require an unusually large disk, but a formation within a disk closer to the primary followed by outward migration (from dynamical interactions) would be possible. The low-mass substellar companions studied here have a mass representing only 0.75--4\% of the mass of their primaries. 
If these companions actually formed like stars, then it would imply that the fragmentation process can produce objects having only about 1\% of the mass of the primary star. In any case, further observations of these systems using high-contrast imaging techniques and radial velocity to search for additional companions would be useful to help understand their origin. For example, if they formed within a disk and were ejected outward, then a more massive object would likely reside in the system at a much smaller separation.

\newpage
\acknowledgments
\noindent F.-R. Lachapelle is supported by a Research Fellowship from the {\it Fonds de Recherche du Qu\'ebec -- Nature et Technologies}. 
DL is supported in part through grants from the Natural Sciences and Engineering Research Council, Canada (NSERC), and from the Universit\'e de Montr\'eal. 
Additional support for this work came from NSERC grants to RJ. 
ChH acknowledges an ERC starting grand under the FP7 program of the European Union. 
The authors wish to thank \'Etienne Artigau for his help with some aspects of this work.
We would also like to thank the anonymous referee for a thorough and useful review of our manuscript. 
Based on observations obtained at the Gemini Observatory, which is operated by the Association of Universities for Research in Astronomy, Inc., under a cooperative agreement with the NSF on behalf of the Gemini partnership: the National Science Foundation (United States), the National Research Council (Canada), CONICYT (Chile), the Australian Research Council (Australia), Minist\'{e}rio da Ci\^{e}ncia, Tecnologia e Inova\c{c}\~{a}o (Brazil) and Ministerio de Ciencia, Tecnolog\'{i}a e Innovaci\'{o}n Productiva (Argentina).

\bibliographystyle{apj}

\end{document}